\documentclass[%
 reprint,%
 aip,amssymb, amsmath,%
 jap%
]{revtex4-1}

\usepackage{docs}%
\usepackage{bm}%
\usepackage[colorlinks=true,linkcolor=blue]{hyperref}%
\expandafter\ifx\csname package@font\endcsname\relax\else
 \expandafter\expandafter
 \expandafter\usepackage
 \expandafter\expandafter
 \expandafter{\csname package@font\endcsname}%
\fi
\usepackage{graphicx}%
\usepackage{dcolumn}%
\usepackage{bm}%

\usepackage[caption=false]{subfig}
\usepackage{multirow} 
\usepackage{rotating} 
\usepackage{longtable}

\graphicspath{{./img/}}

\begin{document}

\title{Microwave dual-mode resonators for coherent spin-photon coupling}%

\author{C. Bonizzoni}
\email[Mail to:]{claudio.bonizzoni@unimore.it}
\affiliation{Universit\'a di Modena e Reggio Emilia - Dipartimento di Scienze Fisiche, Informatiche e Matematiche, via G. campi 213/A, 41125 Modena, Italy}
\affiliation{CNR Nano Istituto Nanoscience - sezione S3, via G. campi 213/A, 41125 Modena, Italy }

\author{F. Troiani}
\affiliation{CNR Nano Istituto Nanoscience - sezione S3, via G. campi 213/A, 41125 Modena, Italy }

\author{A. Ghirri}
\affiliation{CNR Nano Istituto Nanoscience - sezione S3, via G. campi 213/A, 41125 Modena, Italy }

\author{M. Affronte}
\affiliation{Universit\'a di Modena e Reggio Emilia - Dipartimento di Scienze Fisiche, Informatiche e Matematiche, via G. campi 213/A, 41125 Modena, Italy}
\affiliation{CNR Nano Istituto Nanoscience - sezione S3, via G. campi 213/A, 41125 Modena, Italy }

\date{\today}%
\revised{unknown 201x}%


\begin{abstract}

We implement superconducting YBCO planar resonators with two fundamental modes for circuit quantum electrodynamics experiments. We first demonstrate good tunability in the resonant microwave frequencies and in their interplay as it emerges from the dependence of the transmission spectra on the device geometry. We then investigate the magnetic coupling of the resonant modes with bulk samples of DPPH organic radical spins. The transmission spectroscopies performed at low temperature show that the coherent spin-photon coupling regime with the spin ensembles can be achieved by each of the resonator modes. The analysis of the results within the framework of the Input-Output formalism and by means of entropic measures demonstrates coherent mixing of the degrees of freedom corresponding to two remote spin ensembles and, with a suitable choice of the geometry, the approaching of a regime with spin-induced mixing of the two photon modes.  

\end{abstract}

\maketitle

\section{\label{sec.intr}Introduction}

Circuit Quantum Electrodynamics (circuit QED) experiments have recently reached the multimode strong coupling regime, in which single qubits are simultaneously coupled to a large number of discrete, spatially separated and non degenerate photonic modes of a cavity \cite{McKayPRL15,SundaresanPRX15,BosmanNPJQuantInf17,NaikNatCommun17,vaidyaPRX2018}. Multiple hybridization of collective spin degrees of freedom provided by spin ensembles with single photonic modes has been also demonstrated \cite{zhangNatComm2015,ghirriPRA2016,astnerPRL2017}. In all such kinds of experiments a crucial role is played by the possibility of tailoring the composition of the coherent  hybridization involving multiple microwave photonic modes and spin ensembles, even if spatially separated. This is one of the main requirements for the implementation of schemes for quantum information processing \cite{meierPRB2004, DongJModOpt09} or in proposing new circuit QED architectures including auxiliary modes \cite{leekPRL2010,troianiARXIV2018} or lines \cite{kubogrezesPRL2011,jenkinsDaltonTrans2016} to separate qubit state readout and manipulation from photon storage and coherent operations. \\
In this work we investigate the mode hybridization resulting from the coherent coupling between two spin ensembles of  concentrated 2,2-diphenyl-1-picrylhydrazyl (DPPH) organic radical powders and the resonant modes of planar superconducting dual-mode patch resonators (DMRs) \cite{lancaster2006,zhuIEEEMicrowWireCompLett2005,zhuIEEEtransMicrowTech1999,cassineseIEEEtransApplSuperc2001} at microwave (MW) frequencies. These devices display two fundamental resonant modes due to the introduction of symmetry-breaking defects, which lift their degeneracy \cite{zhuIEEEMicrowWireCompLett2005,lancaster2006,najiSCIREP2017,zhuIEEEtransMicrowTech1999} and allow the achievement of a good tunability of their energy separation. Although DMRs find application as filters or antennas in telecommunications, to our knowledge, they have not been explored yet in the field of circuit QED. Here we focus on superconducting YBCO/Sapphire DMRs with different dimensions and coupling geometry and we characterize the transmission of the system in the framework of the Input-Output formalism. We derive and characterize the composition of the hybridized modes in terms of photonic and spin components as a function of the applied magnetic field and in terms of entropic measures of the system. We finally show the achievement of a high degree of multiple coherent spin-photon mixing.\\

\subsection{\label{sec.DMR}Cross Slotted Dual Mode Resonators (DMRs)}

\begin{figure}[ptb]
\centering
 \begin{minipage}{0.47\columnwidth}
 \includegraphics[width=\columnwidth]{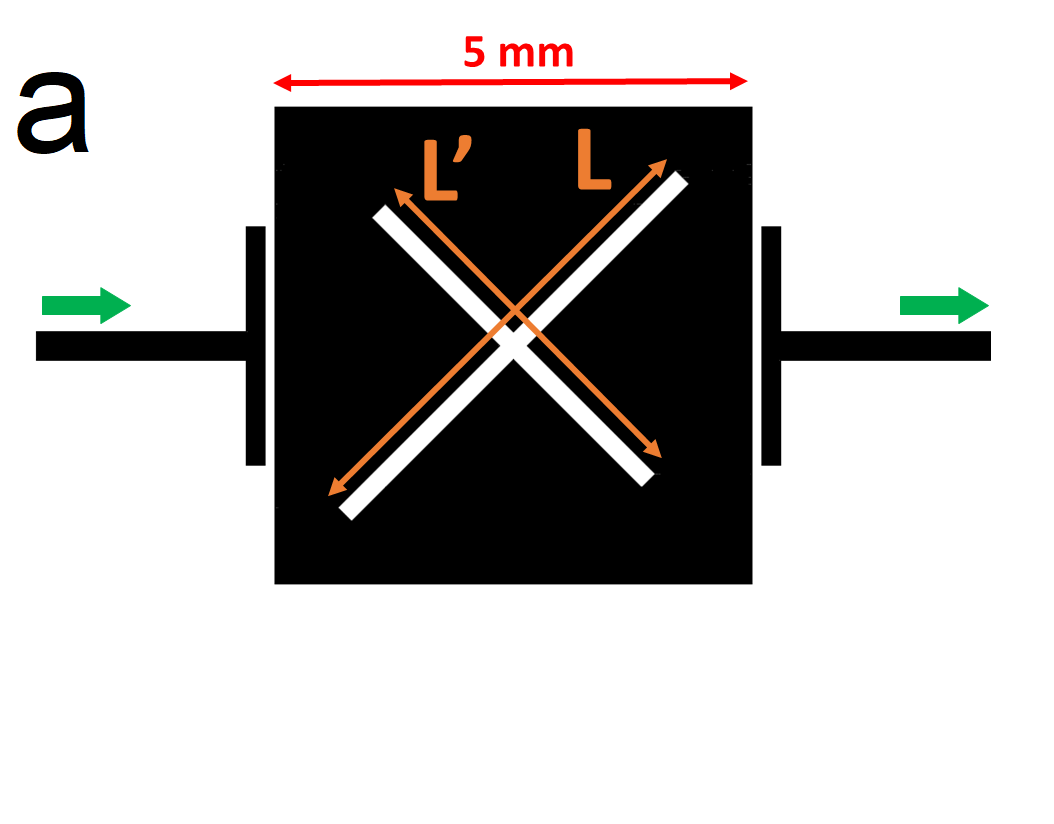}
 \end{minipage}
 \begin{minipage}{0.48\columnwidth}
 \includegraphics[width=\columnwidth]{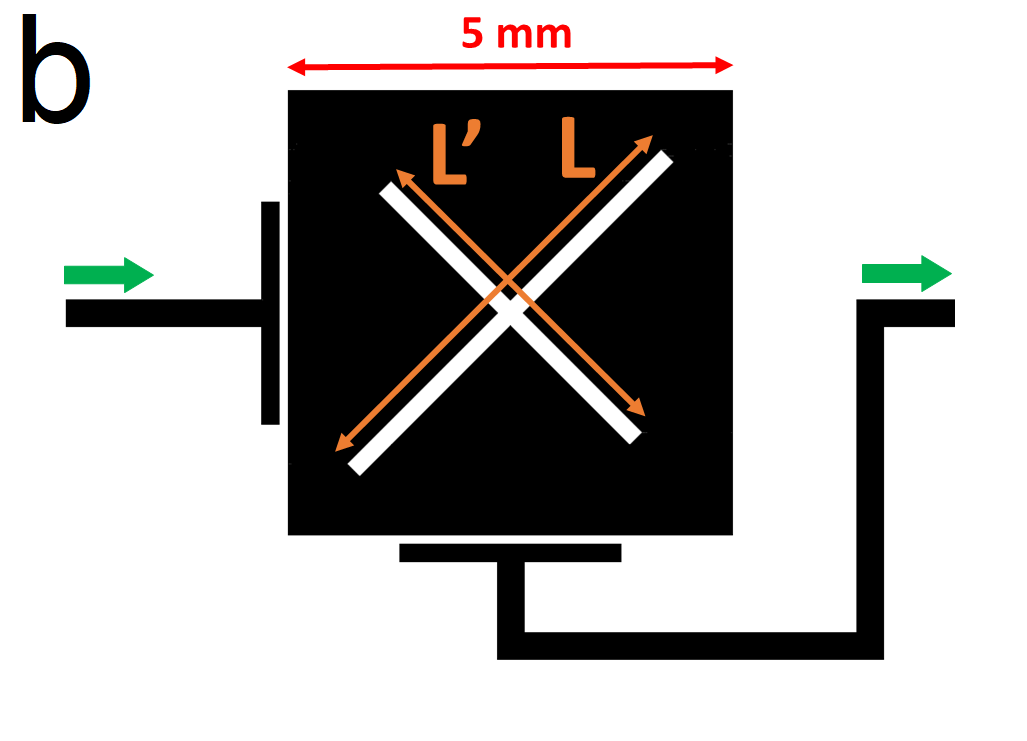}
 \end{minipage}
\centering
 \caption{\label{Fig00} Sketch of the DMRs with parallel (a) and the perpendicular (b) coupler configuration. Green arrows represent the path of the microwave signal through the devices. The lengths of longer slot is $L=5\,$mm, while $L^{'}$ is takes the values 4,4.4, 4.5, 4.7 mm.}
\end{figure}

The DMRs under investigation are shown in Fig. \ref{Fig00}. The device consists of a central square patch ($5\times 5\,$mm$^{2}$), with two perpendicular intersecting slots ($200\text{ }\mu$m width) of lengths $L$ and $L'$. Due to the resulting X shape, these kind of devices are ofter referred to as \emph{Cross Slotted} DMRs, and the symmetry breaking is provided by the asymmetry in the length of the slots. The design of the device has been essentially adapted from the ones reported in Ref. \cite{zhuIEEEtransMicrowTech1999,lancaster2006} to obtain fundamental modes at 7-8 GHz on a Sapphire substrate. The microwave signal (green arrows) is injected and collected through two capacitive coupling gaps (100 $\mu$m), which are fed by two on-chip microstrip transmission lines. These are arranged in two different configurations, hereafter referred to as {\it parallel} (a) and {\it perpendicular} (b). Due to the capacitative gaps, the coupling between the DMR and its feed line is driven by the MW electric field \cite{zhuIEEEtransMicrowTech1999}. The square patch, the feed lines, and the couplers are made of $330\,$nm thick high-$T_c$ superconducting films of Yttrium Barium Copper Oxide (YBa$_{2}$Cu${_3}$O${_7}$, YBCO for short) on Sapphire \cite{ESI}. The initial commercial $10 \times 10\,mm^{2}$ YBCO/Sapphire films are purchased from Ceraco and then patterned by means of optical lithography followed by wet chemical etching in diluted Phosphoric acid, according to the procedure previously reported in Ref. \cite{ghirriAPL2015_notce}.

\section{\label{sec.theory}Theoretical Model}

	\subsection{Hamiltonian for DMRs coupled to spin ensembles}

\begin{figure}[h]
 \includegraphics[width=0.98\columnwidth]{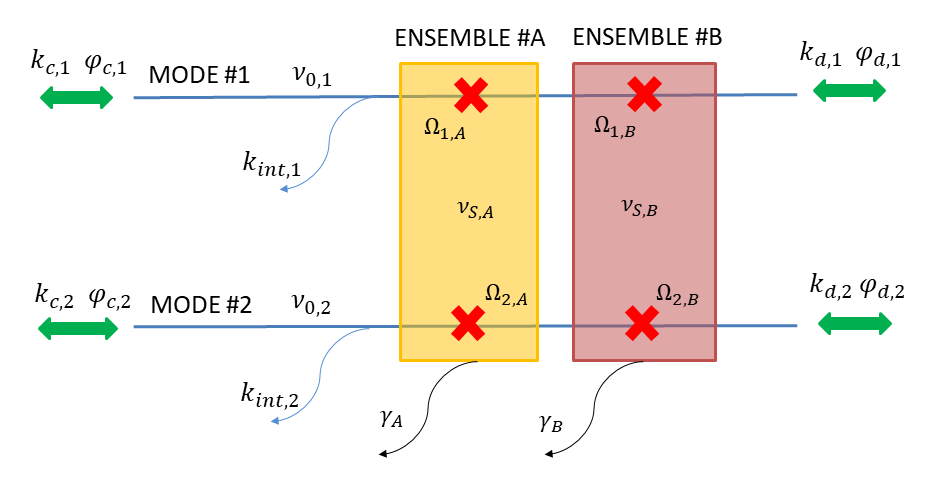}%
 \caption{\label{fig.model}
Pictorial representation of a DMR coupled to two spin ensembles. The two (photon) modes \#1 and \#2 are characterized by their frequencies $\nu_{0,l}$ ($l=1,2$), their external ($k_{\alpha,l}$), with $\alpha = c,d$ the two sides of the device) and internal ($\kappa_{int,l}$) decay rates. The external couplings are also characterized by the phase $\varphi_{\alpha,l}$. We take for simplicity $\varphi_{c,1}=\varphi_{c,2}=\varphi_{d,2}=0$, while $\varphi_{d,1}=\phi$ assumes different values in the parallel and perpendicular geometries (Sec. \cite{ESI}). The two spin modes $\chi=A,B$ correspond to the ensembles \#A and \#B respectively, and they have excitation energies $h\nu_{s,\chi}$ and HWHM linewidths $\gamma_{\chi}$. Each of them can couple to each of the resonant modes through its corresponding coupling rate $\Omega_{l,\chi}$.}
\end{figure}

We model the DMRs coupled to the two spin ensembles as in Fig. \ref{fig.model}. We consider $N_c=2$ independent harmonic oscillators with frequencies $\nu_{0,l}$ ($l=1,2$), corresponding to the two resonant (photon) modes of the DMRs. Each mode can couple to $N_s=2$ effective spins corresponding to the localized spin ensemble. In the weak-excitation regime (where the fraction of excited spins is much smaller than 1, {\it i.e.} when the mean photon number $n_{ph,l}$ is much lower with respect to the number of effective spins $N_{eff,\chi}$), one can treat the spin degrees of freedom in terms of bosonic operators \cite{holsteinPhysRev1940}, and thus introduce an additional harmonic oscillator for each of the $N_s$ spin ensembles. The resulting Hamiltonian is:
\begin{eqnarray}
\label{eq.io_hamilt}
H=   
\sum_{l=1}^{N_c} h \nu_{0,l} a_{l}^{\dagger}a_{l} 
&+& 
\sum_{\chi=1}^{N_s} h \nu_{s,\chi} b_{\chi}^\dagger b_{\chi}
\nonumber\\
&+& 
\sum_{l=1}^{N_c} \sum_{\chi=1}^{N_s} h \Omega_{l,\chi} (a_{l}^{\dagger}b_{\chi} + b_{\chi}^\dagger a_{l})  .
\end{eqnarray}
Here, $h \nu_{s,\chi} = g_\chi\mu_B B_l $ is the Zeeman energy of the spins that form the $\chi$-th spin ensemble, with $g_\chi$ and $B_i$ the Land\'e \textit{g}-factor and the resonant magnetic field with Mode $l$, respectively. $b_\chi$ ($b_\chi^{\dagger}$) is the annihilation (creation) operator corresponding to the $\chi$-th spin mode. The interaction between the spin ensembles and the cavity modes is accounted by the last term. In the following we will refer to $\Omega_{1,A},\Omega_{2,B}$ as \emph{direct coupling rates}, since they quantify the coupling between each mode and its corresponding ensemble, and to $\Omega_{1,B},\Omega_{2,A}$ as \emph{cross coupling rates}, since they describe the interaction with each mode and the ensemble localized on the other one.  The collective spin-photon coupling rate $\Omega_{l,\chi}$ depends by the effective number of spins $N_{eff,\chi}$ in the ensemble and by the single spin coupling rate, $\Omega_{S_{l,\chi}}$, according to \cite{jenkinsJOP2013,abeAPL2011,ghirriAPL2015_notce}:
\begin{equation}
\Omega_{l,\chi}=\Omega_{S_{l,\chi}}\sqrt{N_{eff,l,\chi}}.
\label{eq.coupling_rate}
\end{equation}
The effective number of spins $N_{eff,l,\chi}$ involved in the coupling with a resonant mode depends by the temperature according to \cite{chiorescuPRB2010,jenkinsJOP2013,abeAPL2011, ghirriAPL2015_notce}: 
\begin{math}
N_{eff,l,\chi}=N_{0,l,\chi}P_{l,\chi}(T),
\label{eq.effective_spin_num}
\end{math}
where $N_{0,l,\chi}=\rho V_{eff,l,\chi}$ is the maximum number of spins which can be coupled by the resonant mode, and (for spin $1/2$) can be considered as the zero temperature limit for $N_{eff,l,\chi}$. $N_{0,l,\chi}$ depends by the density of the spins in the ensemble, $\rho$, and by the effective volume $V_{eff,l,\chi}$, \textit{i.e.} the volume of the sample which is effectively interacting with the MW resonant mode. The function $P_{l,\chi}(T)$ (with $0<P_{l,\chi}(T)\leq 1$) is the polarization function of the ensemble and, for DPPH, it's given by the Brillouin law with $J=1/2$\cite{ghirriAPL2015_notce}:
\begin{math}
P_{l,\chi}(T)=tanh\left(\frac{h\nu_{0,l}}{2 k_{B}T}\right). 
\end{math}
Eq. \ref{eq.io_hamilt} can be re-written in a diagonal form in terms of $N_c + N_s$ hybridized spin-photon modes \cite{ghirriPRA2016}:
\begin{equation}
H = \sum_{k=1}^{N_c + N_s} h \mu_{k} c_k^\dagger c_k,
\end{equation}
which holds for the general case of $N_{c}$ photon modes and $N_{s}$ spin modes. Here, each of the $k = 1,\dots, N_c + N_s$ hybridized mode is characterized by its frequency $\mu_{k}$ and by its composition in terms of the bare spin ($b_{\chi}$) and photon ($a_l$) modes, according to the coefficients $\eta^c_{k,l}$ and $\eta^s_{k,\chi}$ of the annihilation operator  
\begin{math}
c_k = \sum_{l=1}^{N_c} \eta^c_{k,l} a_l + \sum_{\chi=A}^{N_s} \eta^s_{k,\chi} b_\chi ,
\end{math} 
whose square moduli are normalized to 1. We will analyze and discuss the square amplitudes of such coefficients ($|\eta^c_{k,l}|^2$ and $|\eta^s_{k,\chi}|^2$) as a function of the static magnetic field in Sec. \ref{sec.weights}.

	\subsection{Entropy and Spin Character for hybridized modes}

The degree of coherent spin-photon mixing can be quantified in terms of entropic measures. In particular, the modal entropy $S_k$ of a $k^{th}$ hybridized mode can be defined as:
\begin{equation}
S_{k} = - \sum_{l=1}^{N_c} |\eta^c_{k,l}|^{2} \ln |\eta^c_{k,l}|^{2} - \sum_{\chi=A}^{N_s} |\eta^s_{k,\chi}|^{2} \ln |\eta^s_{k,\chi}|^{2} .
\label{eq.modal_entropy}
\end{equation}

The entropy in Eq. \ref{eq.modal_entropy} ranges from 0, where the $k-$th mode coincides with a bare spin or photon mode, up to $ln(N_{c}+N_{s})$, when the full mixing between all the bare modes is achieved ($|\eta^c_{k,l}|=|\eta^s_{k,\chi}|=1/\sqrt{N_c+N_s}$). It follows that the entropy resulting from the mixing of only two bare modes can have a maximum value $ln(2)$. Values for the modal entropy $ln(2) < S_{k} \leq ln(3)$ thus denote a coherent mixing involving up to three bare modes. Finally, values of modal entropy $S_{k} > ln(3)$ necessary imply the coherent hybridization of 4 bare modes.\\
Although quantifying the degree of mixing, the modal entropy does not give any information about the type of bare modes involved and about the dominant character of the hybridization (bright = photon or dark = spin). Thus, this is quantified by the spin ($P_{s,k}$) or by the photon character ($P_{c,k}$), \textit{i.e.} by the sums of the probabilities corresponding to the bare spin or photonic modes:
\begin{eqnarray}
P_{s,k} = \sum_{\chi=A}^{N_s} |\eta^s_{k,\chi}|^{2} ,\text{   } 
P_{c,k} = \sum_{l=1}^{N_c} |\eta^c_{k,l}|^{2} = 1 - P_{s,k}.
\label{eq.spin_character}
\end{eqnarray}
Obviously, in the limiting cases of a purely photonic (spin) mode, $P_{s,k}$ equals 0 (1). The entropy and the spin character calculated for the experiments of this work are discussed in Sec. \ref{sec.entropy_spin_char}.

	\subsection{Input-Output formalism for DMRs}

Within the framework of the Input-Output formalism \cite{walls}, we focus on the complex scattering parameter $S_{21}(\nu)$ which represents, for a given probing frequency $\nu$, the transmission from one side of the cavity to the other one. The essential details of the derivation are left to \cite{ESI} and here we only give the final expression: 

\begin{equation}\label{eq.io_dual_modes}
S_{21} 
= 
\frac{
\kappa_{2} A_{11} e^{-i\phi} + \kappa_{1} A_{22}
-\sqrt{\kappa_{1}\kappa_{2}}\,(A_{21}+A_{12}e^{-i\phi})
}{A_{22} A_{11} - A_{21} A_{12}}.
\end{equation}
The quantity $\kappa_{l}$ ($l=1,2$) is the external photon decay rate for mode \#$l$, resulting from the coupling to external feed capacitors and lines, $i$ is the imaginary unit. The coefficients $A_{lm}$ ($l,m=1,2$) are given by:
\begin{eqnarray}
\label{eq.inout_a_coeff}			
A_{ll} & = & \Delta_{c,l} + \frac{1}{2}(\kappa_{l}+\kappa_{l}+\kappa_{int,l})+\sum_{\chi=A,B} \frac{\Omega_{l,\chi}^2}{i\Delta_{s,\chi}+\gamma_\chi} \nonumber
\\ 
A_{lm} & = & \frac{1}{2} \sqrt{\kappa_{m}\kappa_{l}} \left[ 1 +
                                e^{i(-1)^l\phi} \right] 
       + \sum_{\chi=A,B} \frac{\Omega_{l,\chi} \Omega_{m,\chi}}{i\Delta_{s,\chi}+\gamma_\chi},
\end{eqnarray}
where $\Delta_{0,l}\equiv\nu_{0,l}-\nu$, $\Delta_{s,\chi}\equiv\nu_{s,\chi}-\nu$, while $\gamma_\chi$ and $\nu_{s,\chi}$ are the HWHM linewidth and the transition frequency of the spin ensemble \#$\chi$, respectively. Finally, $\kappa_{int,l}$ is the internal decay rate of the cavity mode $l$, and $\phi$ is the relative phase difference between the external couplings of \#1 and \#2. Note that, due the identical dimensions chosen for the input and the output capacitative coupling gaps (see Fig. \ref{Fig00}), the external coupling with the feed lines of each mode $l$ can be considered as symmetric (\textit{i.e.} $\kappa_{c,l}=\kappa_{d,l}=\kappa_{l}$).   

\section{\label{sec.exp_results}Transmission Spectroscopy of DMRs}

\subsection{\label{sec.exp_setup}Experimental Set-up}

\begin{figure}[h]
\centering
\begin{minipage}{0.39\columnwidth}
 \includegraphics[width=\columnwidth]{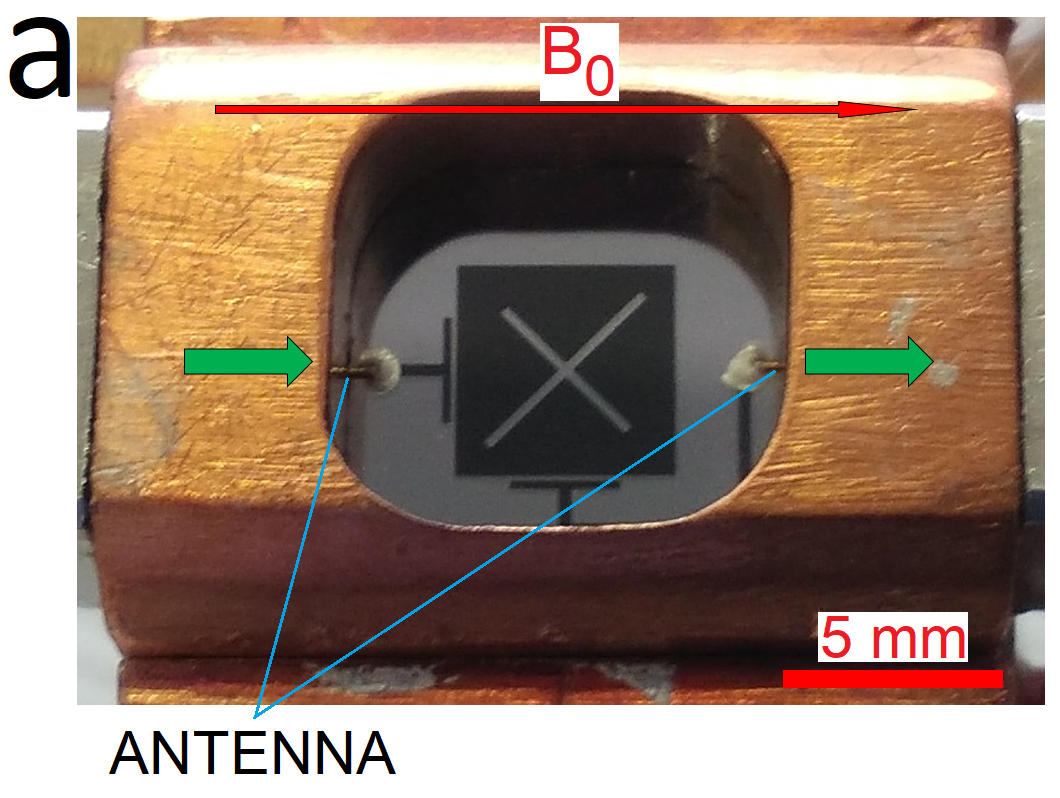}
\end{minipage}
\begin{minipage}{0.55\columnwidth}
 \includegraphics[width=\columnwidth]{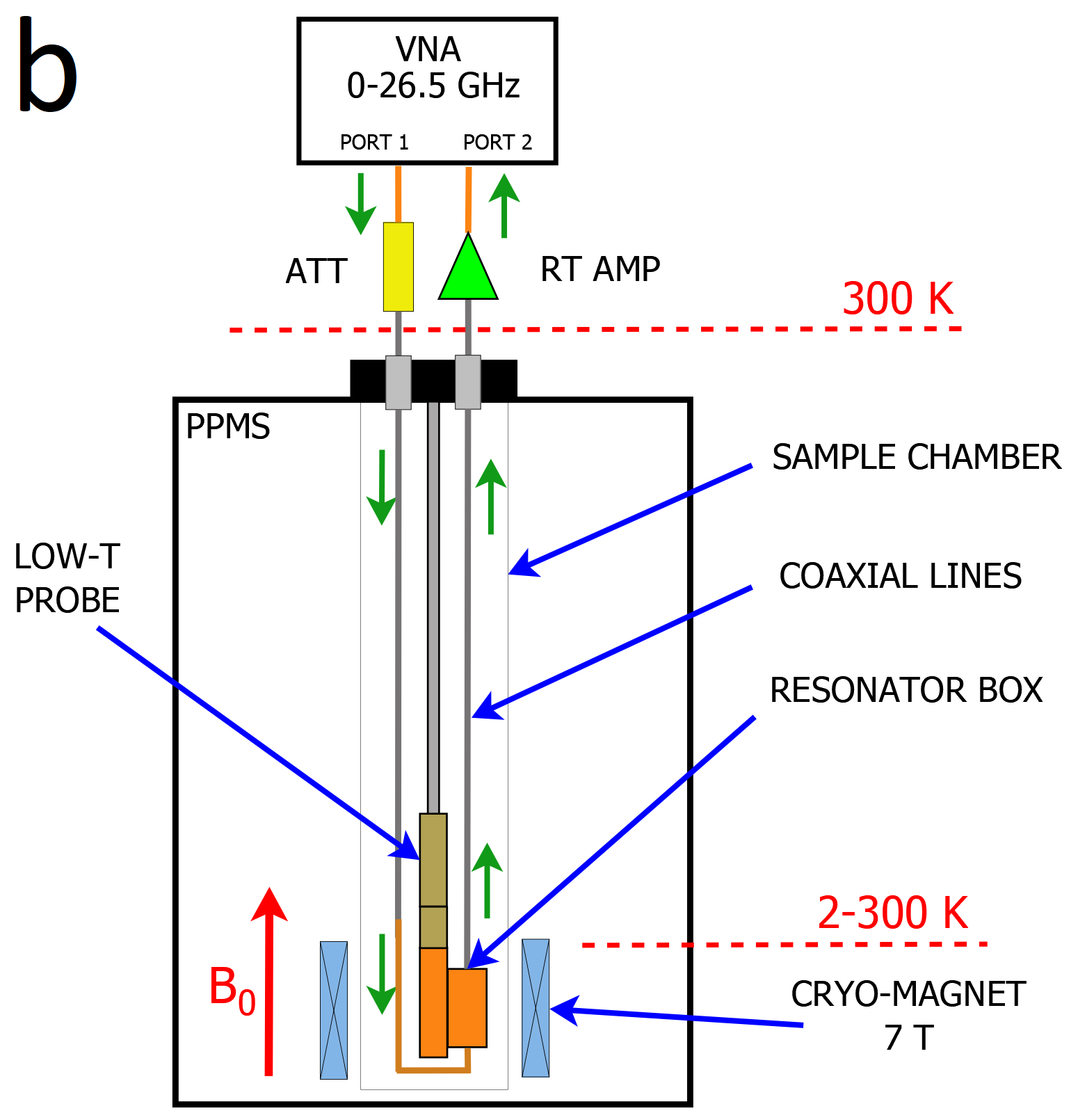}
\end{minipage}
 \caption{\label{fig.exp_setup} (a) Photo of the shielding box with a perpendicular coupled DMR mounted inside. Green arrows show the path of the MW signal through the box, while the red arrow shows the direction of the static magnetic field $B_{0}$. (b) Sketch of the experimental set-up used in this work. Figure (b) is reproduced under Creative Commons Attribution License from Ref. \cite{bonizzoniAdvPhys2018}. 
}
\end{figure}
	
The experimental set-up used is essentially the one previously described in Refs. \cite{bonizzoniAdvPhys2018,bonizzoniDALTON2016} and its reported in Fig. \ref{fig.exp_setup} for better clarity. The DMRs are mounted inside a OFHC Copper shielding box, equipped with two SMA connectors. These are ended with two launching antennas, which are glued on the on-chip transmission lines by means of Silver Epoxy Paste. The box is mounted on a home-made low-T probe with two MW coaxial lines, which is inserted inside a commercial Physical Properties Measurement System (PPMS, 2-300 K, 0-7 T) by Quantum Design. The static magnetic field always lies on the plane of the device, parallel to the input on-chip line. The MW signal is generated and collected with a Vector Network Analyzer (VNA). Additional, optional room temperature attenuators and amplifier can be installed on the input or on the output line respectively, in order to change the input power $P_{in}$. In this work this value refers to the power at the input on-chip transmission line of the DMRs, which is calculated taking into account all the attenuation and the losses along the MW line.  	

\subsection{\label{sec.zero_field}Spectra of bare DMRs}

\begin{figure}[ptb]
\centering
 \includegraphics[width=0.7\columnwidth]{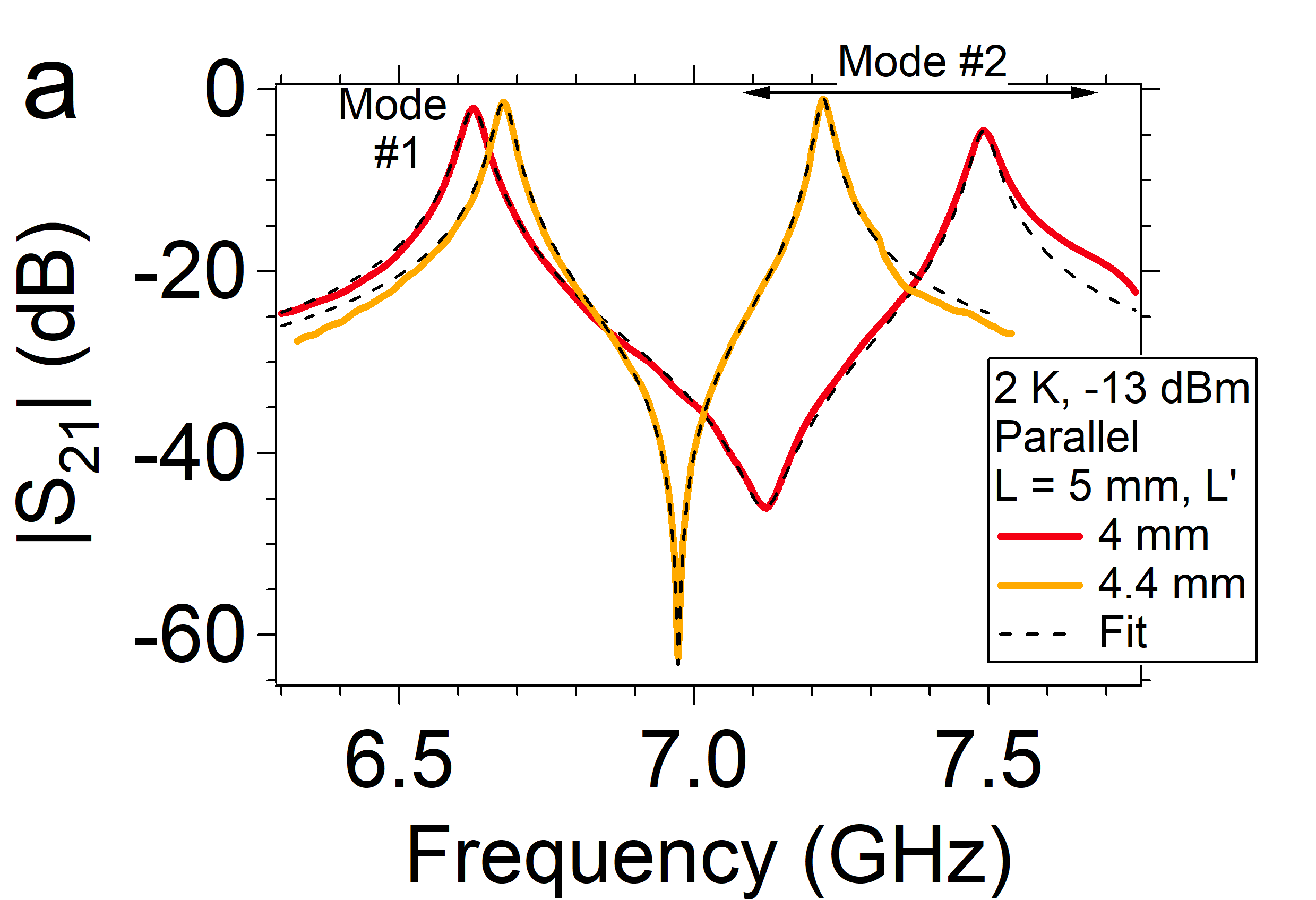}
 \includegraphics[width=0.7\columnwidth]{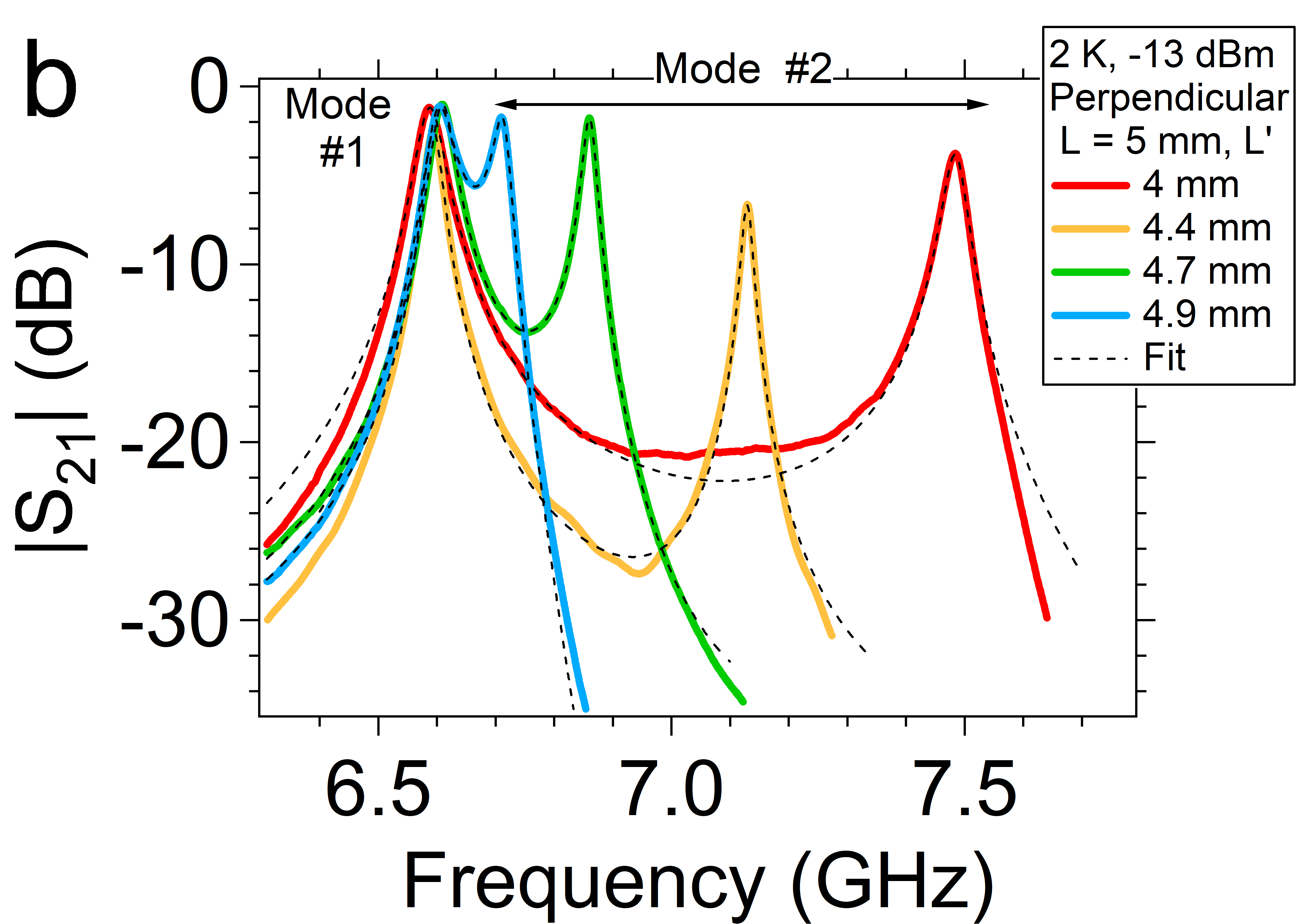}
 \caption{\label{Fig01} Transmission spectra for the empty parallel (a) and the perpendicular (b) coupler configuration. The measures are performed at 2 K, $P_{in}=-13\,$dBm, zero magnetic field and for different lengths of the shorter slot ($L^{'}=4,4.4, 4.5, 4.7$ mm, see legends). Dashed lines correspond to the simulated spectra based on Eq. \ref{eq.io_dual_modes}.}
\end{figure}

We first consider the case of \emph{bare} DMRs, in which the devices are not loaded with spin ensembles. Figure \ref{Fig01} shows the transmission scattering parameter $|S_{21}|$ \cite{pozar} measured in the parallel and in the perpendicular configuration, at 2 K and $P_{in}=-13\,$dBm, for different values of $L^{'}$. The spectra are characterized by the presence of a microwave photon mode \#1, with frequency $\nu_{0,1}\approx6.6\,$GHz, essentially independent on $L^{'}$, and of a mode \#2, whose frequency $\nu_{0,2}$ decreases monotonically with $L^{'}$. In this way a wide tunability of the frequency difference $\Delta\nu=\nu_{0,2}-\nu_{0,1}$ can be achieved, ranging between $100$ and $900\,$MHz in our devices. 
The transmission spectra obtained in the parallel configuration are characterized by a clear dip between the two peaks which, on the contrary, is absent in the perpendicular configuration (Fig. \ref{Fig01}.a). Input-Output simulations (Eq. \ref{eq.io_dual_modes}) of the transmission spectra and electromagnetic simulation of the field distributions \cite{ESI} show that this difference follows from the different phase values at the capacitive couplers in the two configurations. The field simulations also reveal that the electric and magnetic components of the resonant modes are respectively perpendicular and parallel to the plane of the device. Besides, the maxima of the amplitude of the magnetic field are located at the edges of each slot, and the mode with lower (higher) frequency is mainly localized on the longer (shorter) slot. Furthermore, we note that the use of YBCO makes the mode frequency and quality factor stable with respect to temperature and applied magnetic fields (see \cite{ESI}), as previously observed for planar single-mode resonators \cite{ghirriAPL2015_notce}.

We use Eq. \ref{eq.io_dual_modes} to simulate the zero field transmission spectra in order to estimate the main physical parameters for the empty DMR. Here the phase $\phi$ is fixed to the value given by the electromagnetic simulations (see \cite{ESI}), while all the coupling rates are fixed at $\Omega_{l,\chi}=0$. We obtain the values for the internal and the external decay rates and of the resonant frequency of each mode (Table \ref{Table.empty_dev}).

\begin{table}[b]
\centering
\begin{ruledtabular}
\begin{tabular}{c|cccc|cc} 
                         &  \multicolumn{4}{c}{Perpendicular}                           & \multicolumn{2}{c}{Parallel} \\
\hline 
$L^{'}\text{ (mm)}$      & 4             &  4.4           & 4.7          &    4.9       & 4             & 4.4         \\ 	
\hline 
$\nu_{0,1}\text{ (GHz)}$ &     6.587     &  6.594         & 6.608        &   6.667      &  6.700        & 6.677       \\  
$\kappa_{1}\text{ (MHz)}$ &     22.0      &  12            & 18.2         &  16.6        &   31.0        & 14.1        \\  
$\kappa_{int,1}\text{ (MHz)}$ &     6.5       &  8             & 4.3          &  4.6         &  1.0          & 4.9         \\ 
$\nu_{0,2}\text{ (GHz)}$ &     7.484     &  7.129         & 6.861        &  6.781       & 7.551         & 7.219        \\
$\kappa_{2}\text{ (MHz)}$ &      13.5     &  11.0          & 10.1         &  8.8         & 43.0           & 11.7        \\ 
$\kappa_{int,2}\text{ (MHz)}$ &     15.0      &  5.0           & 4.4          &  8.5         & 0.6           & 3.0         \\
\end{tabular}
\end{ruledtabular}
\caption{\label{Table.empty_dev}Parameters obtained from the simulation of the transmission spectra of the empty DMR, for different lengths $L^{'}$ and coupling configurations according to Eq. \ref{eq.io_dual_modes}. Here the data refer to 2 K, zero magnetic field and input power $P_{in}=-13$ dBm.}
\end{table}

\subsection{\label{sec.dmr_spins_parallel}Coupling spin ensembles to DMRs in the parallel geometry}

\begin{figure}[ptb]
\centering
\includegraphics[width=0.30\columnwidth]{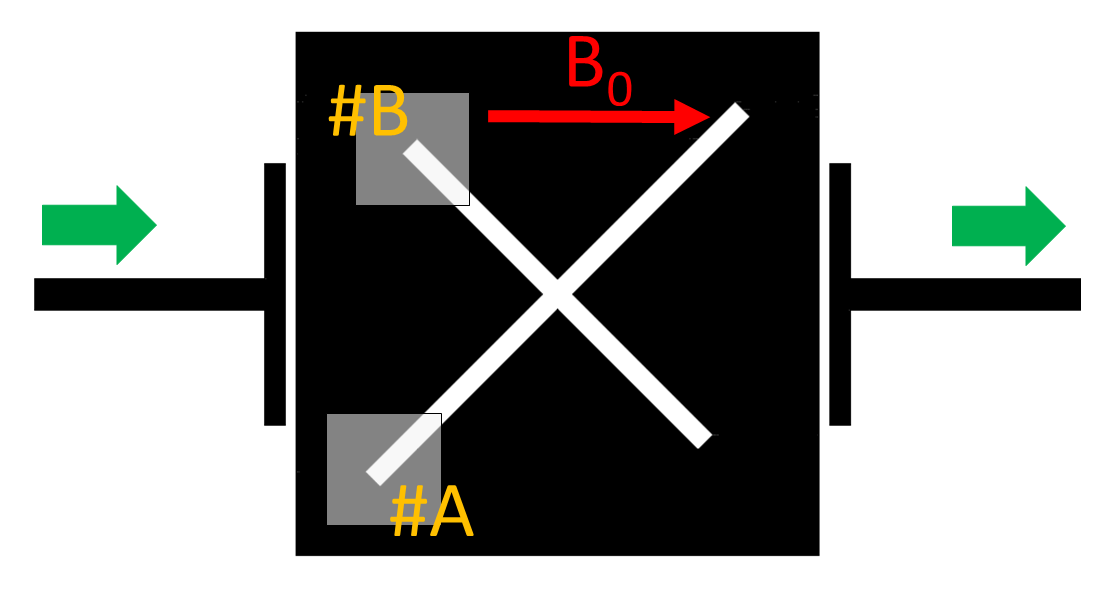}\\ 
\includegraphics[width=0.47\columnwidth]{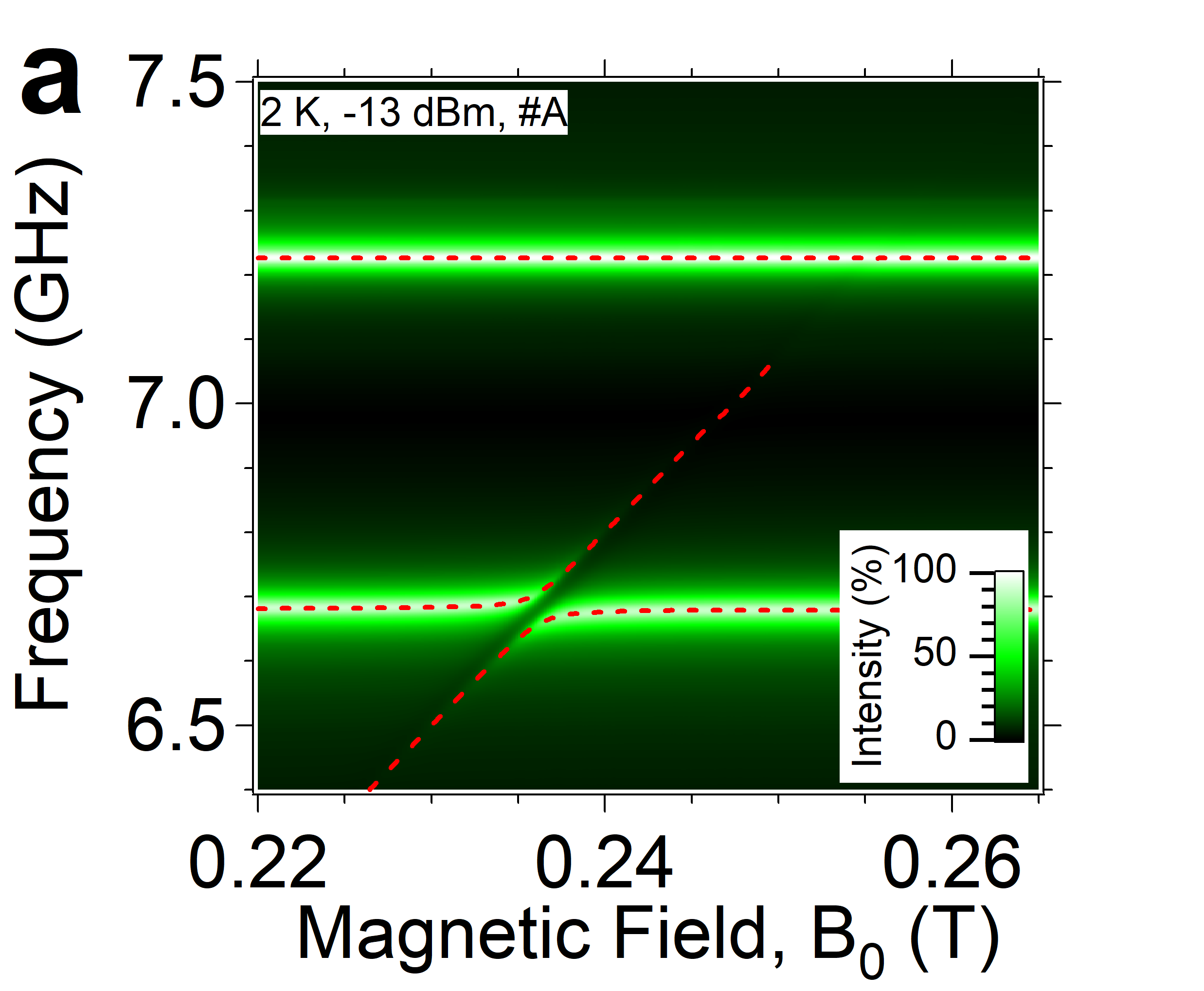} 
\includegraphics[width=0.47\columnwidth]{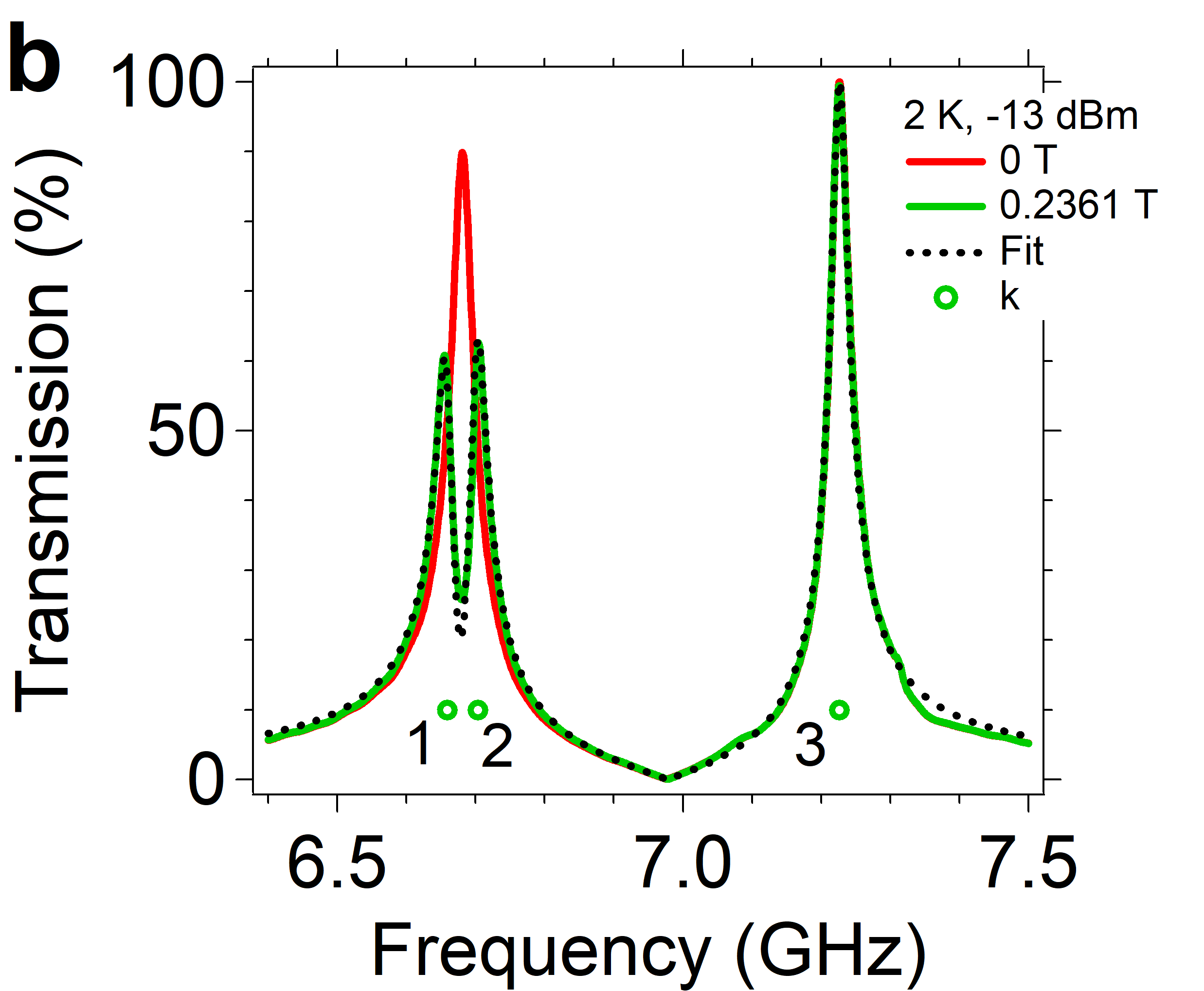}\\ 
\includegraphics[width=0.47\columnwidth]{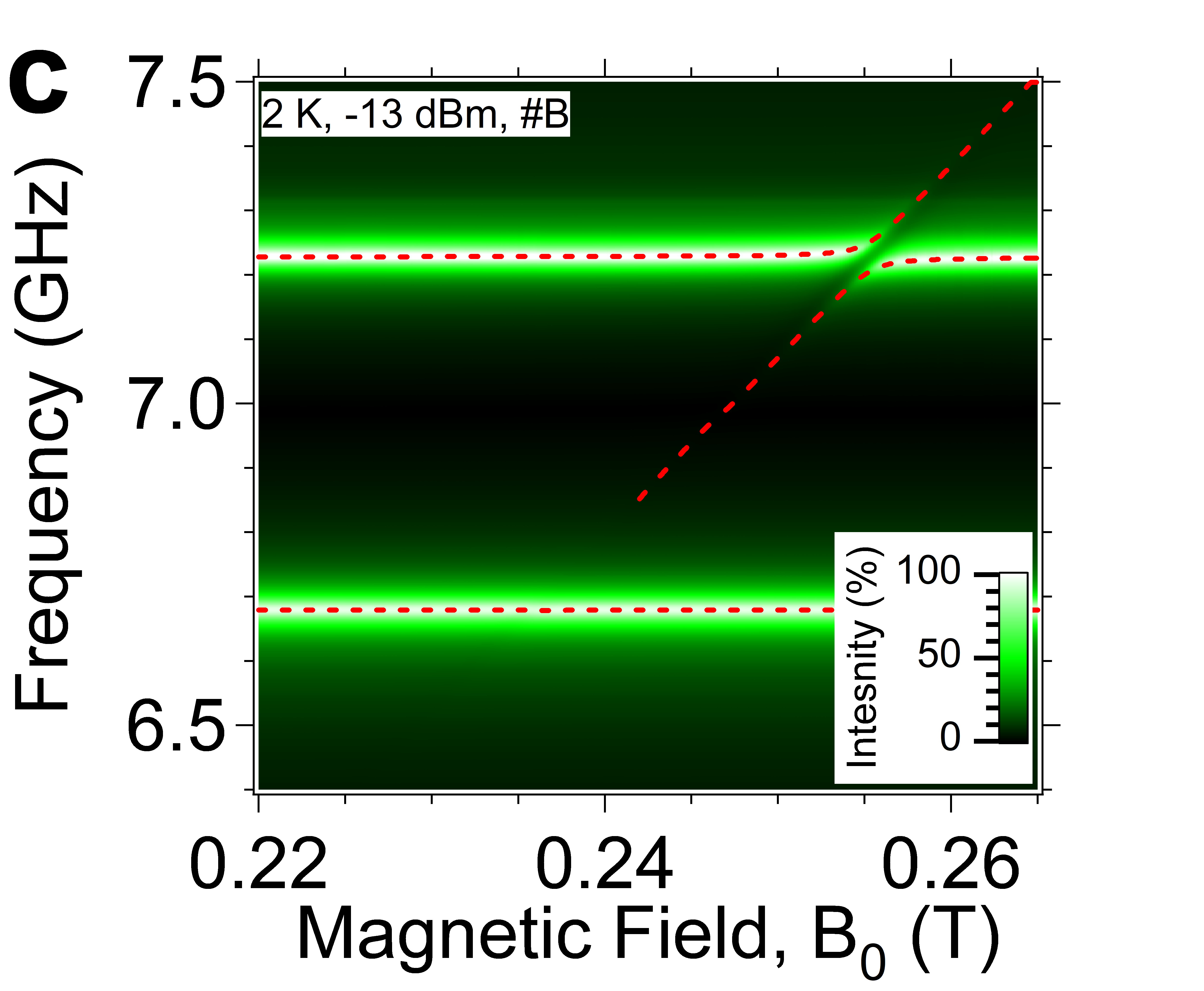} 
\includegraphics[width=0.47\columnwidth]{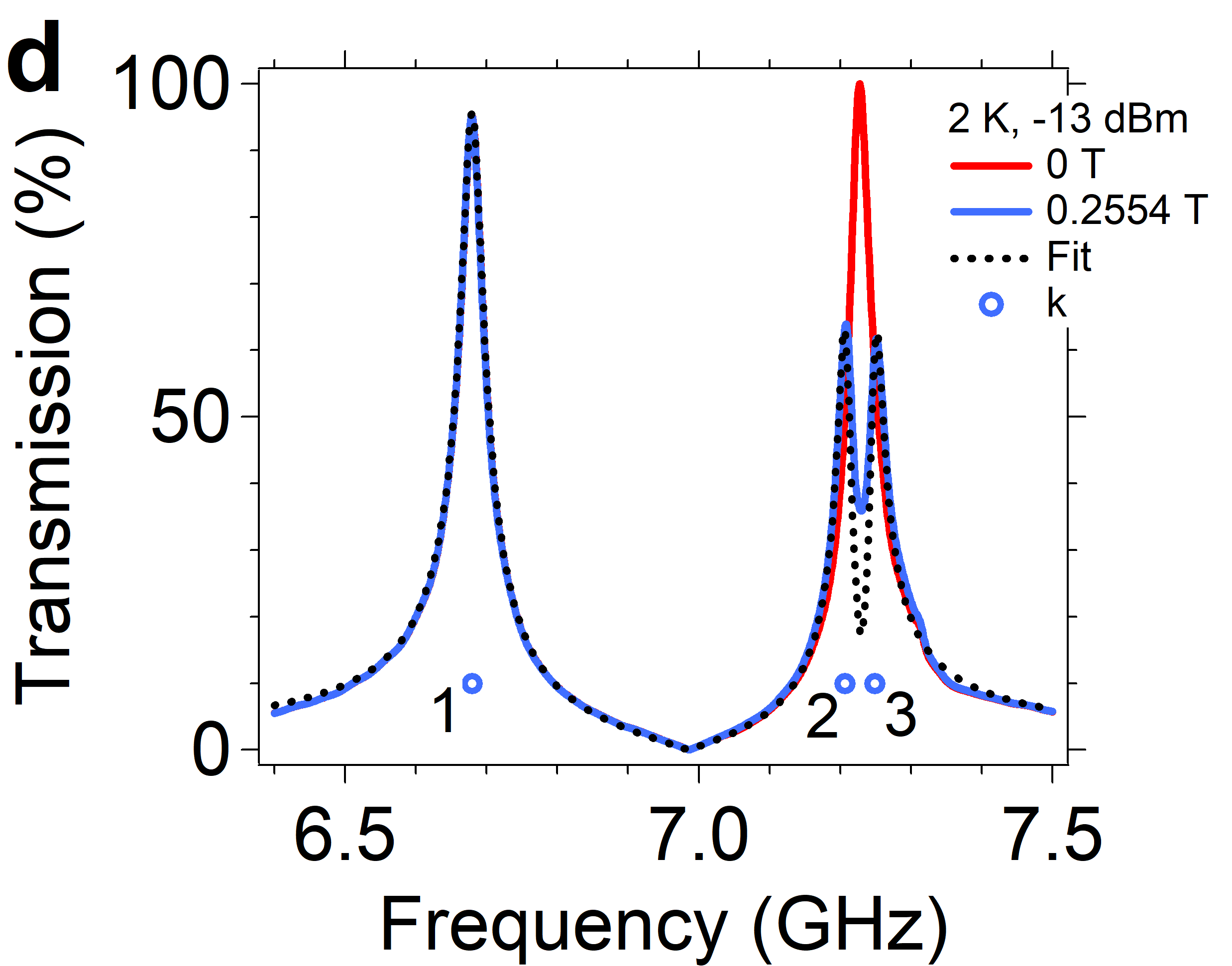}\\ 
\includegraphics[width=0.47\columnwidth]{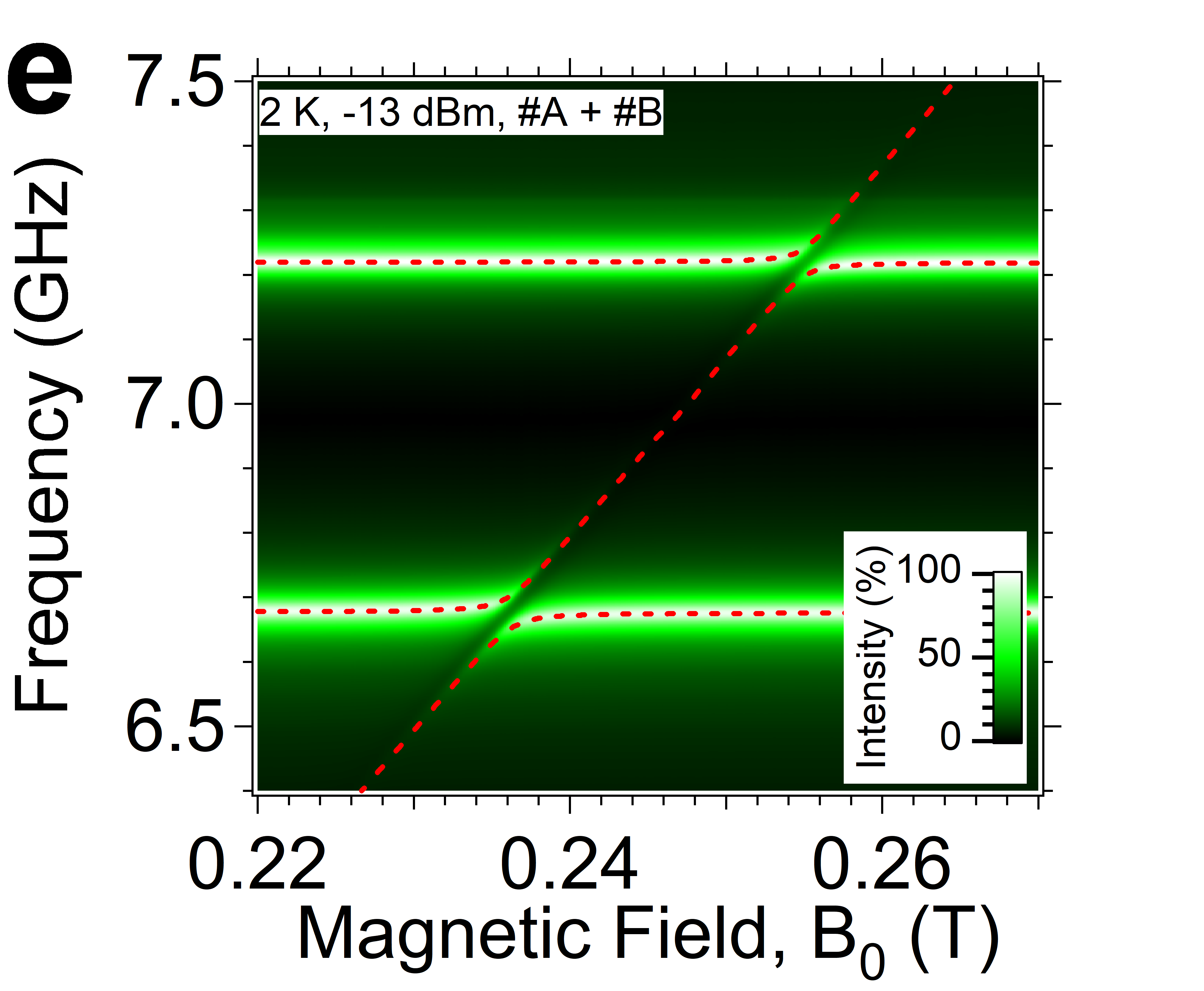} 
\includegraphics[width=0.47\columnwidth]{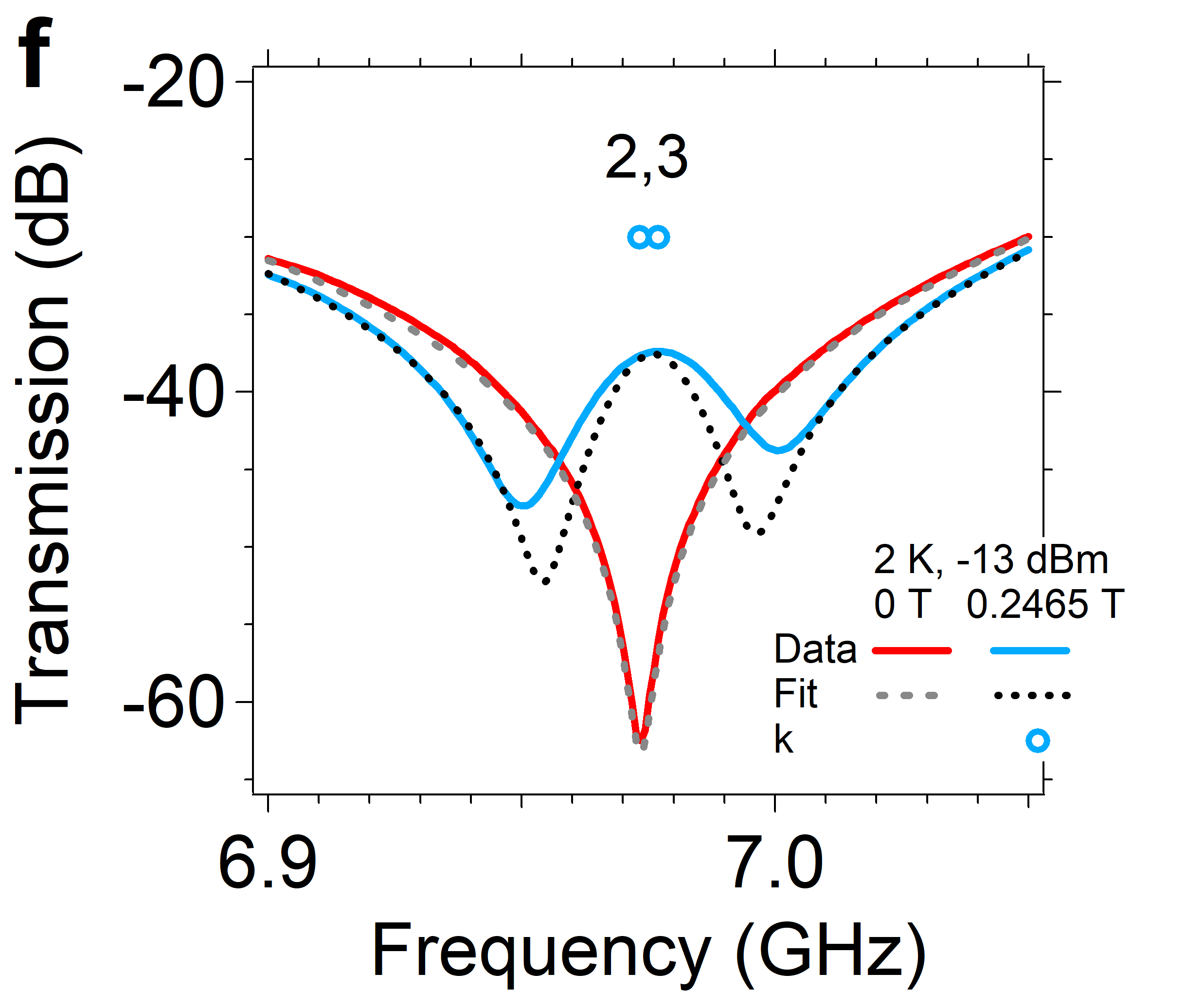} 
\centering
\caption{\label{Fig02}
Transmission spectra obtained in the parallel configuration with $L^{'}=4.4\,$mm. The DMR is loaded either with only ensemble \#A (a,b), only ensemble \#B (c,d) or with both (e,f). The spectra are taken at 2 K and $P_{in}=-13$ dBm. Numbered red dots indicate the computed energies $\mu_{k}$ (in unit of $h$) of the spin-photon mode $k$ at the magnetic field values reported in the legends. Dashed lines are the simulated spectra. The sketch of the DMR on top (same color legend of Fig. \ref{Fig00}) shows the position of the ensembles.}
\end{figure}

We now investigate the magnetic coupling between the photonic modes \#1 and \#2 of the parallel coupled DMRs and two spin ensembles of DPPH. These are hereafter referred to as \#A and \#B, and they are positioned along the slots of length $L$ and $L'$, respectively. Each of the ensembles have dimensions of $\approx 1.2\times 1\times0.9\text{ }$mm$^{3}$, which correspond to an overall number of spins $N_{0,tot}\approx 10^{18}$. We start by considering the DMR in the parallel configuration and with $L^{'}=4.4\,$mm (Fig. \ref{Fig02}), in the case where a single ensemble is placed on the device and positioned to maximize its coupling with the resonant mode. In the presence of ensemble \#A, the transmission spectral map [panel (a)] displays a level anticrossing around the resonant field $B_{0,1}=h\nu_{0,1}/\textit{g}_{A}\mu_{B}$
. The transmission spectra at such field (b) shows a clear Rabi splitting, resulting from the coherent coupling between mode \#1 and the spin ensemble. Similarly, when only ensemble \#B is placed on the resonator, a level anticrossing is observed at the resonant field $B_{0,B}=h\nu_{0,2}/\textit{g}_{B}\mu_{B}$ (c,d). Hence, the coherent coupling regime is achieved for each of the two spin ensembles and for each of the two modes.
When both \#A and \#B are placed on the resonator, the two anticrossings at $B_{0,1}$ and $B_{0,2}$ are still visible in the spectral map (e), while a small peak is observed at the intermediate field $B_{0,3}$ (f), with a height which is approximately equal to the sum of those obtained with \#A or \#B alone.\\ 

\begin{table}[ptb]
\caption{\label{Table1} Estimated values of the main physical parameters, derived from the simulation of the experimental spectra in the parallel configuration at 2 K and $P_{in}=-13\,dBm$. }
\centering
\begin{ruledtabular}
\begin{tabular}{c|ccc} 
Spin ensembles                    & \#A               & \#B            & \#A + \#B   \\ 
\hline
$\Omega_{1,A}\text{ (MHz)}$   & $22\pm1$          & -           & $23\pm1$   \\ 
$\Omega_{1,B}\text{ (MHz)}$   & -                  & $2.1\pm0.3$ & $3.0\pm0.5$    \\ 
$\Omega_{2,A}\text{ (MHz)}$   & $1.9\pm0.4$        &  0    & $3.0\pm0.5$    \\ 
$\Omega_{2,B}\text{ (MHz)}$   & -                  & $21.0\pm1.5$    & $19.5\pm1.0$    \\ 
$\gamma_{A}\text{ (MHz)}$     & $8.8\pm0.7$        & -            & $8.0\pm0.5$    \\ 
$\gamma_{B}\text{ (MHz)}$     & -                   & $8.0\pm1$       & $10.3\pm1$   \\ 
%
\end{tabular}
\end{ruledtabular}
\end{table}

In order to estimate the relevant physical parameters from our experiments, we simulate the spectra with Eq. \ref{eq.io_dual_modes}, and we obtained the estimates given in Table \ref{Table1}. The {\it direct} coupling between the spin ensemble \#A (\#B) and the photon mode \#1 (\#2) ($\Omega_{1,A},\Omega_{2,B}\approx 20\,$MHz) is larger than both the spin ($\gamma_{\chi}\approx 8\,$MHz) decay rates, confirming that the coherent coupling regime is achieved. Besides, two smaller {\it cross} couplings ($\Omega_{1,B},\Omega_{2,A} \approx 2-3\,$MHz) need to be included in the simulations in order to properly reproduce the experimental results.

\subsection{\label{sec.dmr_spins_parallel}Coupling spin ensembles to DMRs in the perpendicular geometry}

\begin{figure}[b]
\centering
\includegraphics[scale=0.15]{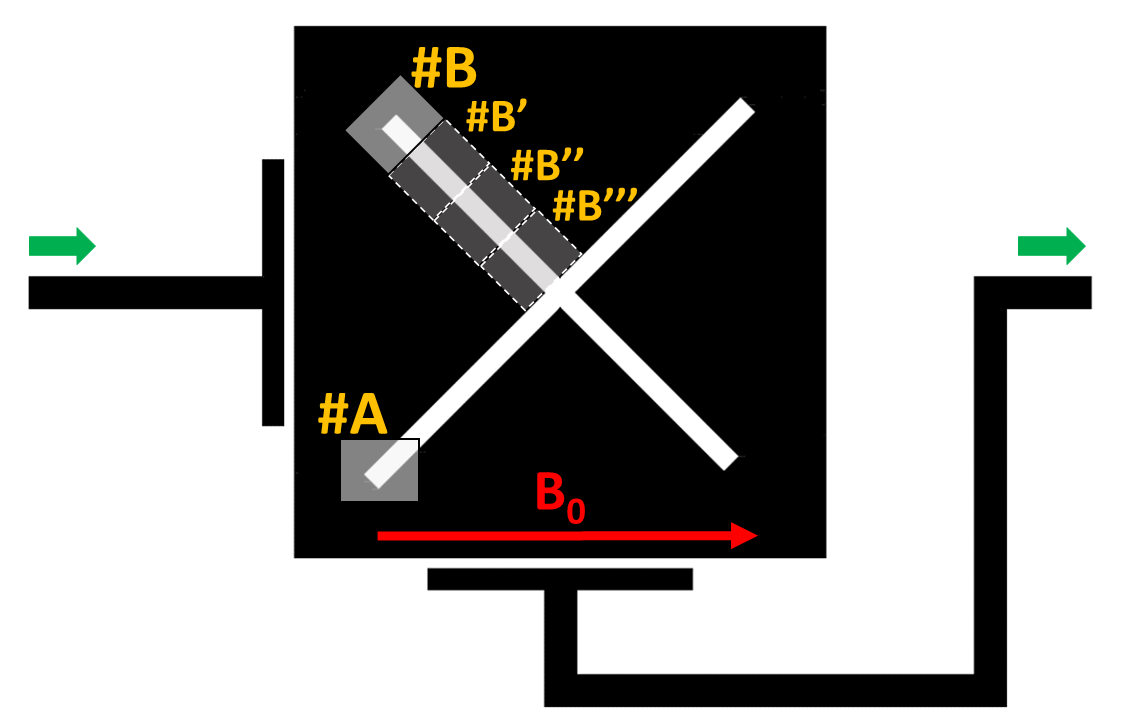}\\
\includegraphics[width=0.48\columnwidth]{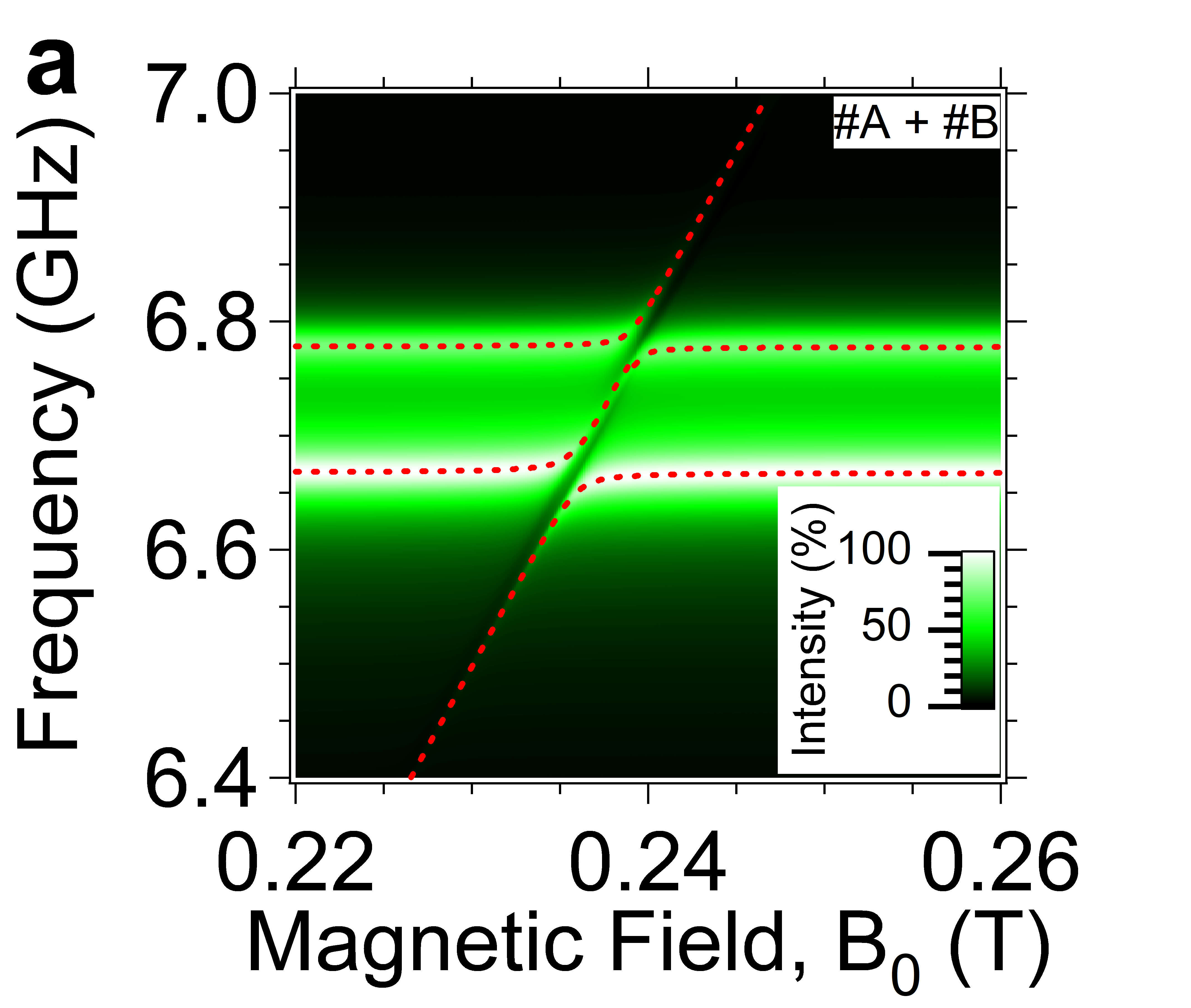} 
\includegraphics[width=0.48\columnwidth]{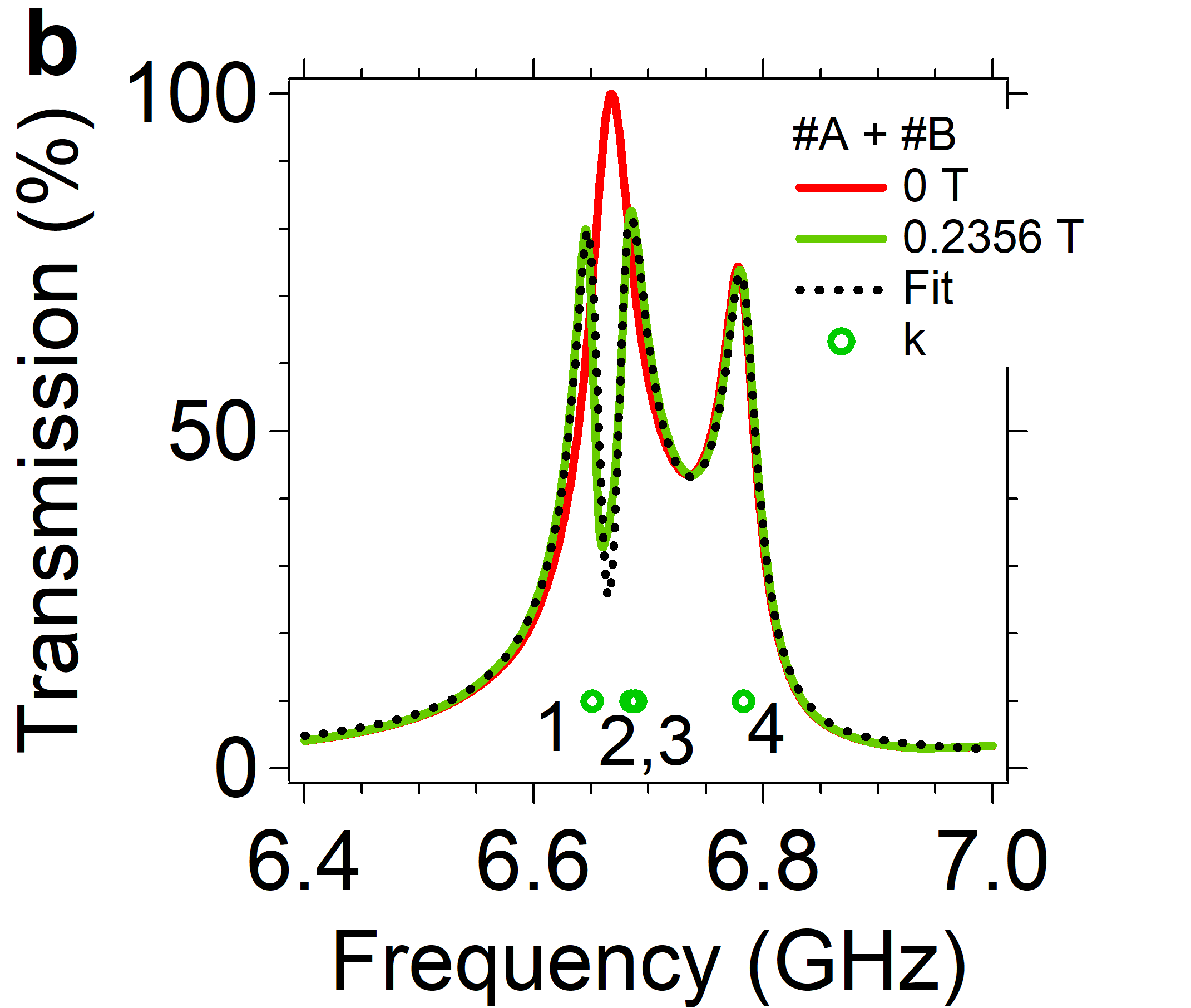}\\
\includegraphics[width=0.48\columnwidth]{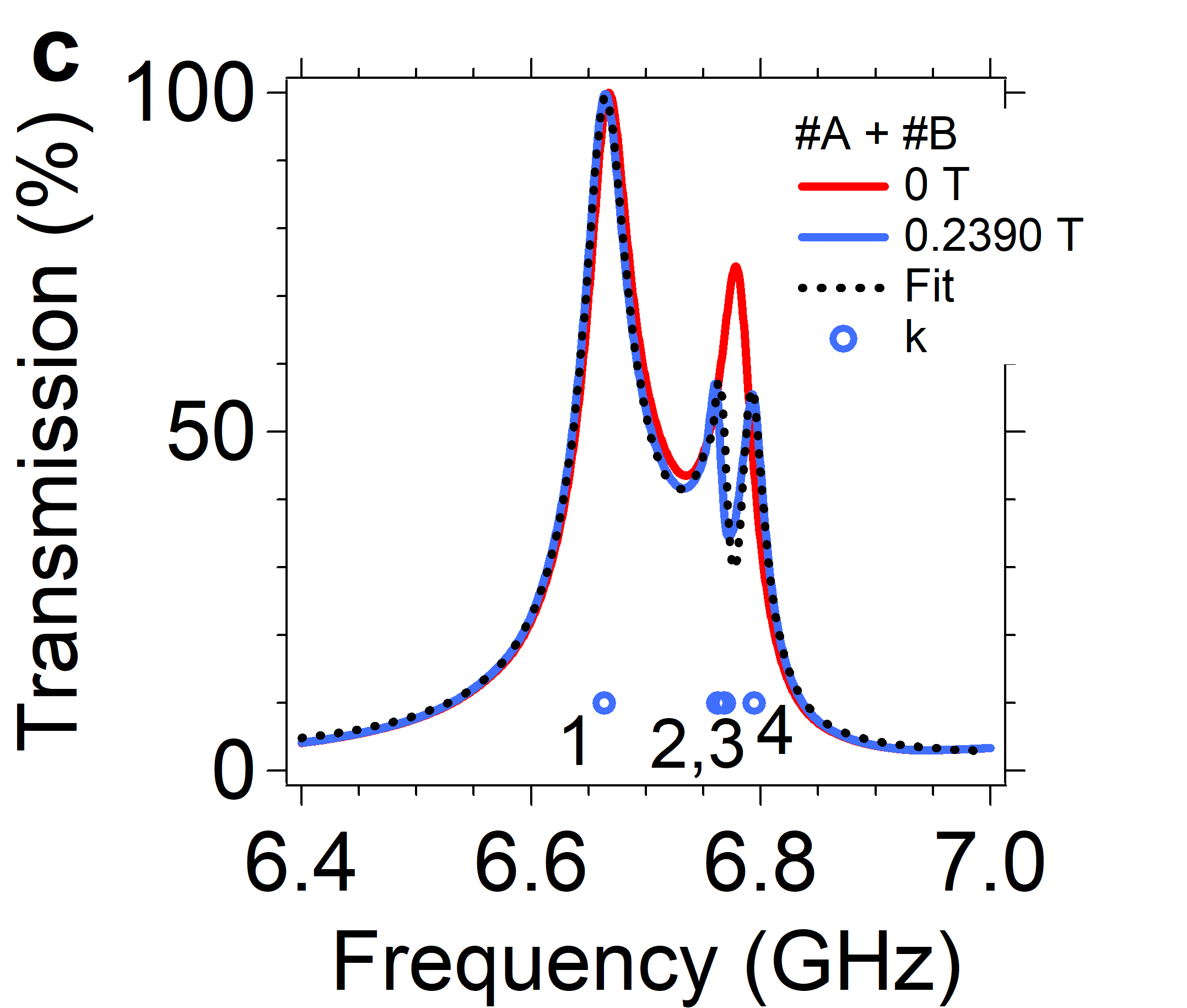}
\includegraphics[width=0.48\columnwidth]{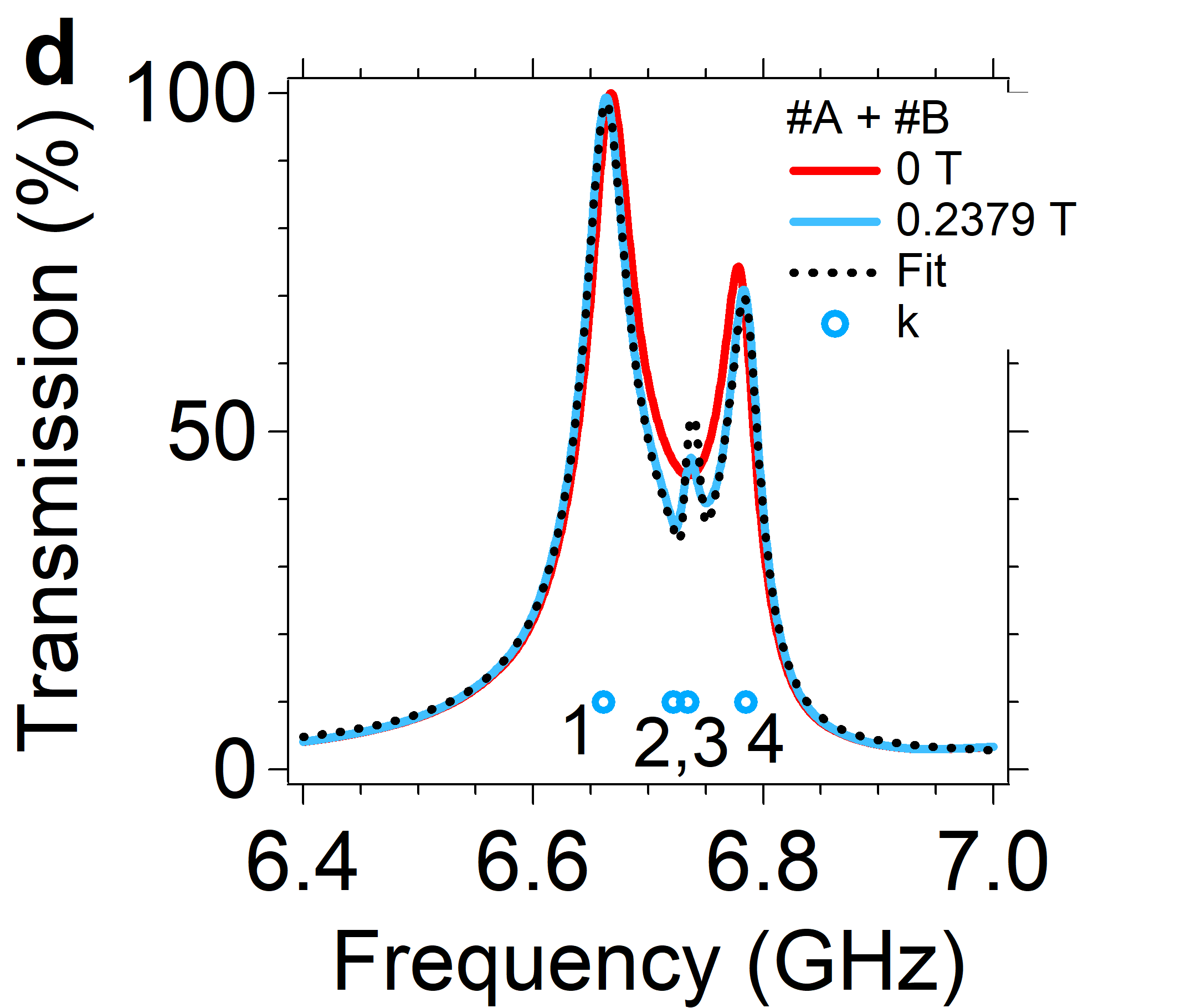}
\centering
\caption{\label{Fig03}Transmission spectroscopy for the geometry of Fig. \ref{Fig01}.b loaded with two ensembles in the \#A and \#B positions at 10 K and $P_{in}=-13$ dBm. The transmission spectra shown are taken at the field $B_{0,1}$ (b), $B_{0,2}$ (c) and at the intermediate field $B_{0,3}$ (d), respectively. Red numbered dots indicate the energies of the hybridized modes $k$ while dashed lines are simulations performed with Eq. \ref{eq.io_dual_modes}. The sketch of the DMR (same symbols and color legend of Fig. \ref{Fig00}) shows the positions of the ensembles along the slots and the orientation of the static magnetic field $B_{0}$.}
\end{figure}

We now consider the case of the perpendicular coupled DMRs and we repeat the spectroscopic characterization described in the previous section. In particular here, choosing a larger value of $L'$ allow us to test the spatial distribution of the modes by changing the position of ensemble \#B along the shorter slot (see \cite{ESI}). The coupling between \#2 and \#B is found to progressively increase as the ensemble is moved from the center to the edge of the slot, as expected from the spatial distribution of the MW magnetic field component (see \cite{ESI}). In the following we focus on the results obtained in the presence of both spin ensembles and with \#B positioned at the edge of the slot $L^{'}$ (Fig. \ref{Fig03}). Two avoided level crossing are visible at the resonant fields $B_{0,1}$ and $B_{0,2}$, while an additional peak is observed at the intermediate field value $B_{0,3}$ (d). The parameters extracted from the comparison between the experimental and the simulated spectra (dashed lines) are reported in Table \ref{Table2} (additional datasets are given in the \cite{ESI}). The values of $\Omega_{1,A}$ and $\gamma_{A}$ are similar to those obtained in the parallel configuration, and don't change as a function of the position of \#B, which affects instead $\Omega_{2,B}$. The effective number of spins estimated for the experiments of this work are reported in \cite{ESI}. Overall at 2 K, the direct coupling between each of the ensembles and its corresponding resonant modes involves $N_{eff,l,\chi}\approx10^{15}$ spins, while for the indirect ones the effective number of spin are $N_{eff,l,\chi}\approx10^{14}$. Note that the condition $N_{eff,l,\chi}\gg n_{ph,l}$, being $n_{ph,l}$ the mean photon number for mode $l$ (see \cite{ESI}), is always met in our experiments. Thus our spin ensembles are always in the low-excitation regime \cite{chiorescuPRB2010,holsteinPhysRev1940}.  

\begin{table}[ptb]
\caption{\label{Table2}Estimated values of the main physical parameters, derived from the simulation of the experimental spectra in the perpendicular configuration at 2 K and $P_{in}=-13\,dBm$.}
\centering
\begin{ruledtabular}
\begin{tabular}{c|ccc} 
Position                    &   \#A + \#B    &  \#A + \#B$^{''}$ & \#A + \#B$^{'''}$ \\ 
\hline 		
$\Omega_{1,A}\text{ (MHz)}$ &   $22.5\pm0.5$   & $22.0\pm2.0$   & $22.5\pm2.0$ \\ 
$\Omega_{1,B}\text{ (MHz)}$ &   $1.2\pm0.4$   & $1.7\pm0.5$     &  $1.5\pm0.3$     \\ 
$\Omega_{2,A}\text{ (MHz)}$ &   $1.0\pm0.3$   & $1.0\pm0.3$     & $1.2\pm0.3$    \\ 
$\Omega_{2,B}\text{ (MHz)}$ &  $15.0\pm 1.5$ &  $10.0\pm1.2$    & $4.5\pm0.5$   \\ 
$\gamma_{A}\text{ (MHz)}$   &  $7.5\pm0.5$    & $7.5\pm0.5$  & $7.5\pm0.5$  \\ 
$\gamma_{B}\text{ (MHz)}$   &   $5.8\pm0.4$     &  $7.8\pm0.5$ & $9.0\pm0.5$  \\ 
\end{tabular}
\end{ruledtabular}
\end{table}

\section{\label{sec.discussion}Mode Hybridization in DMRs}

\subsection{\label{sec.weights}Weighs of the spin and photon components}

\begin{figure}[ptb]
\centering
\includegraphics[width=0.48\columnwidth]{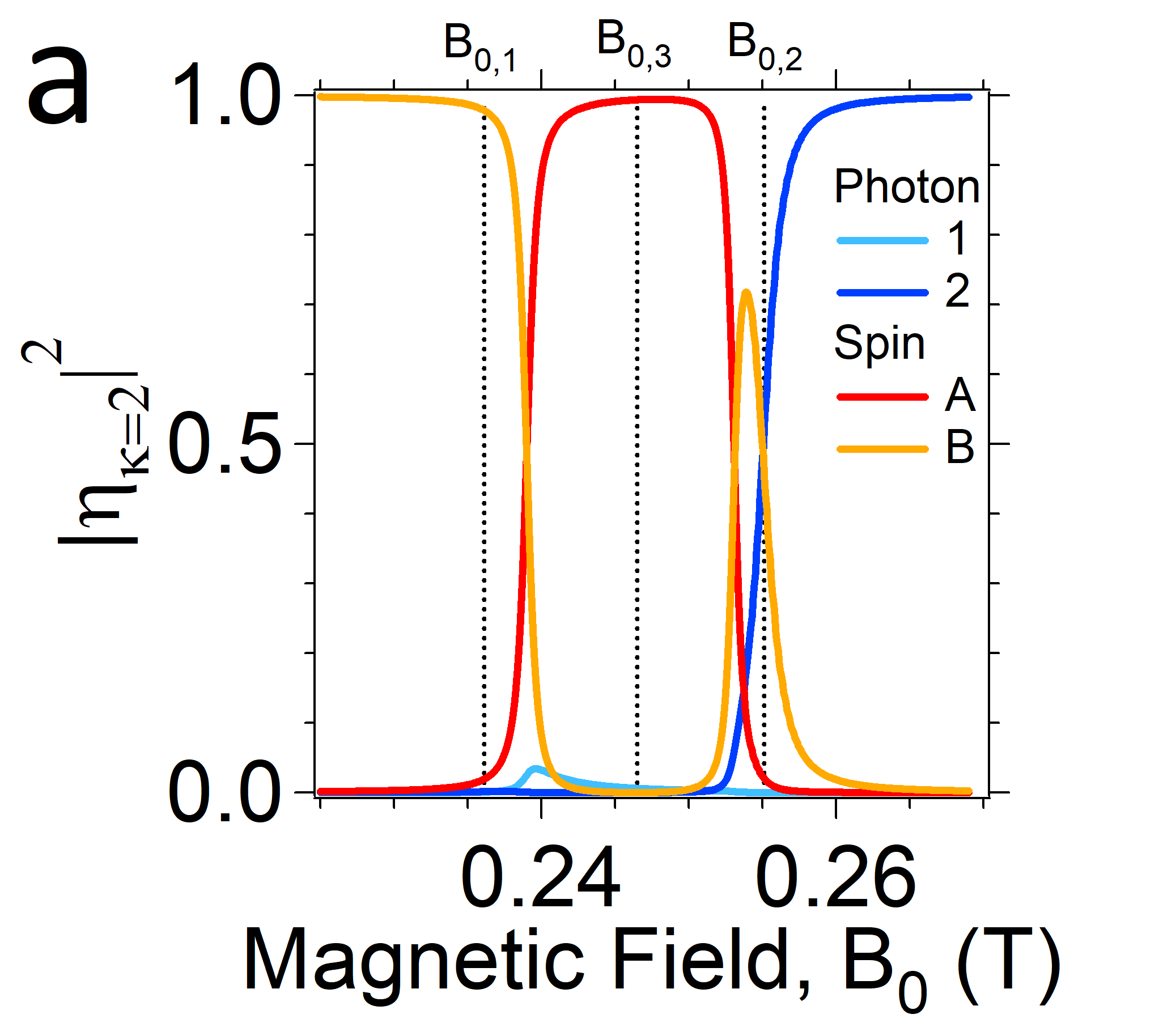}
\includegraphics[width=0.48\columnwidth]{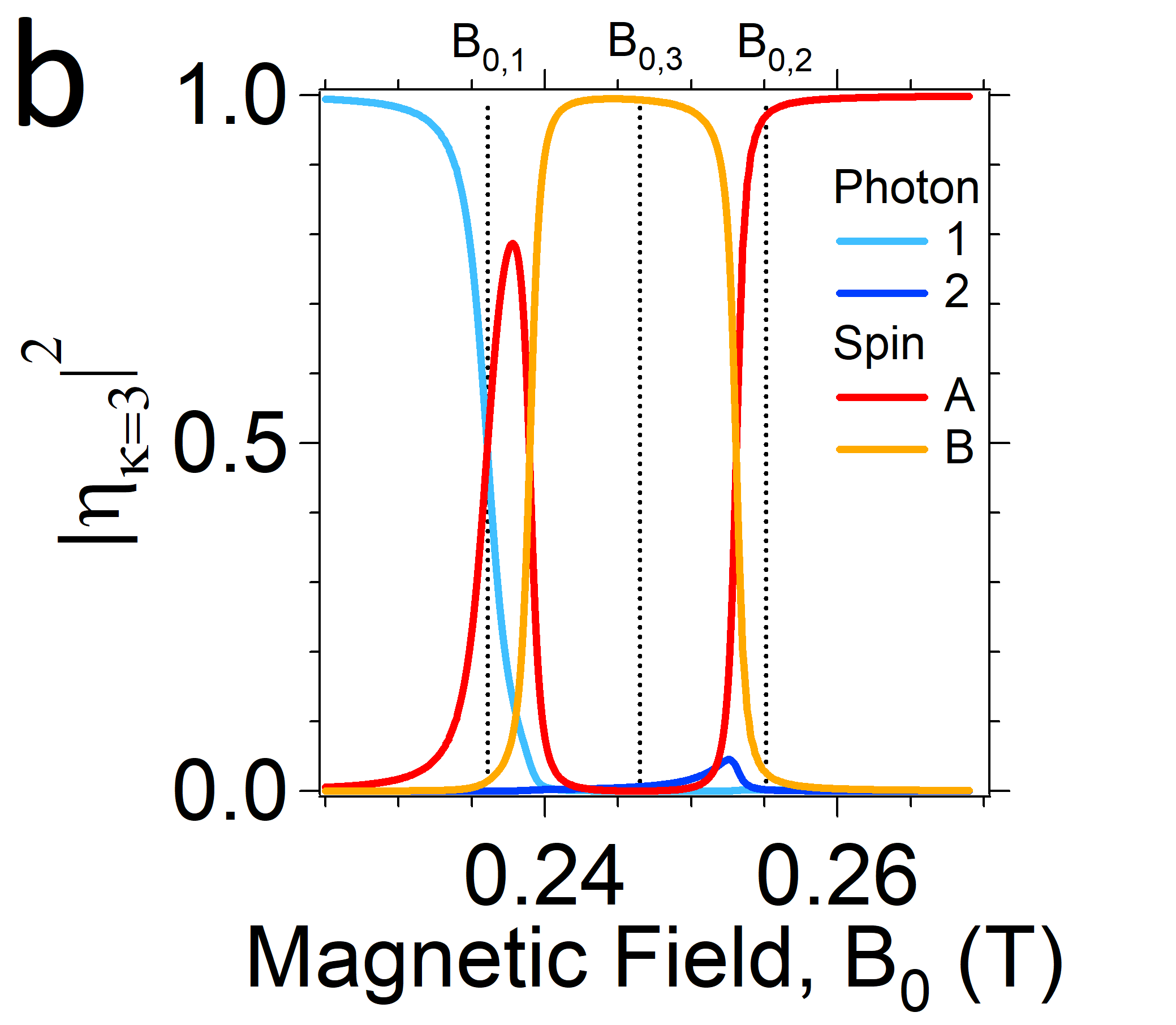} 
\includegraphics[width=0.49\columnwidth]{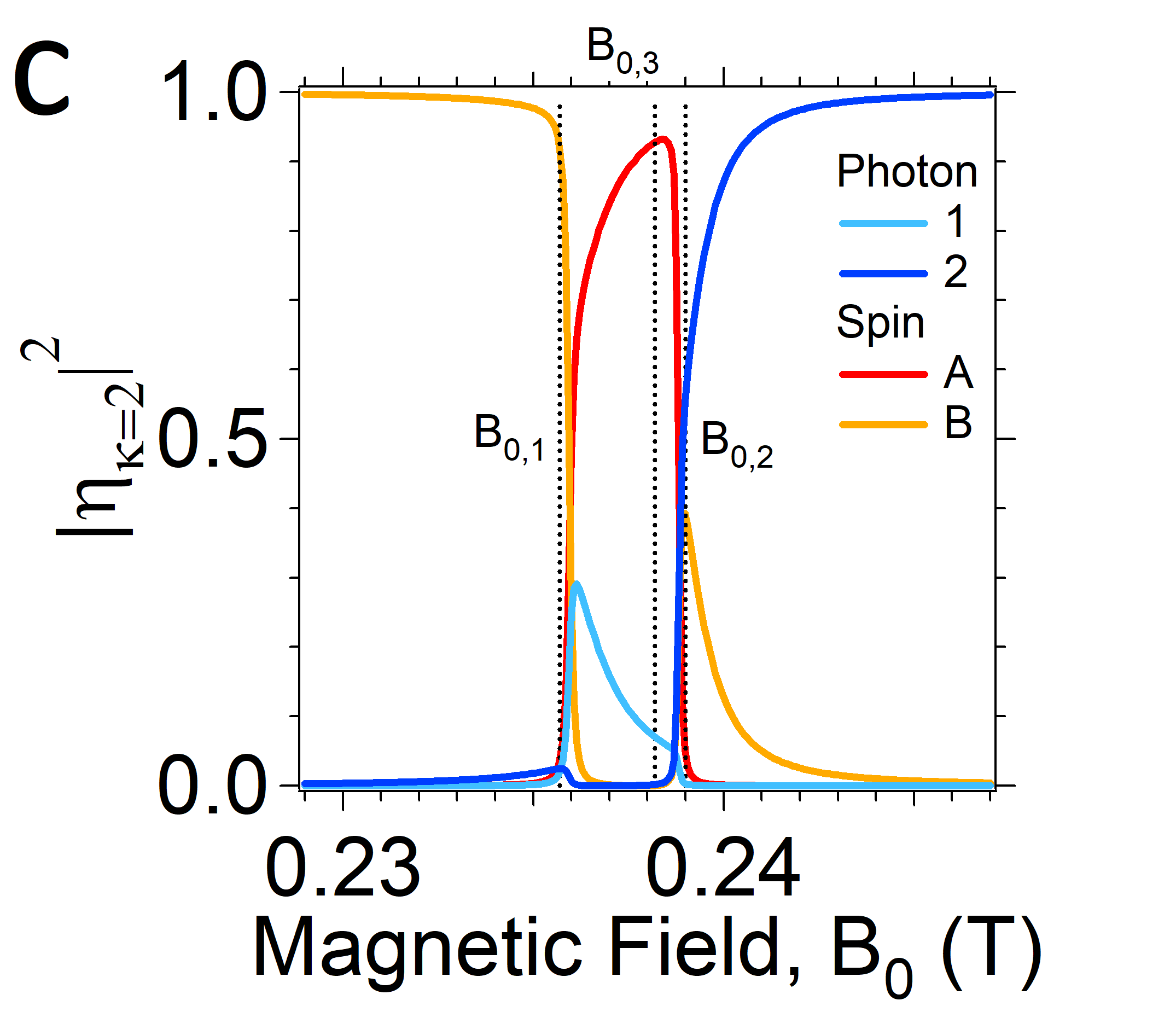}
\includegraphics[width=0.49\columnwidth]{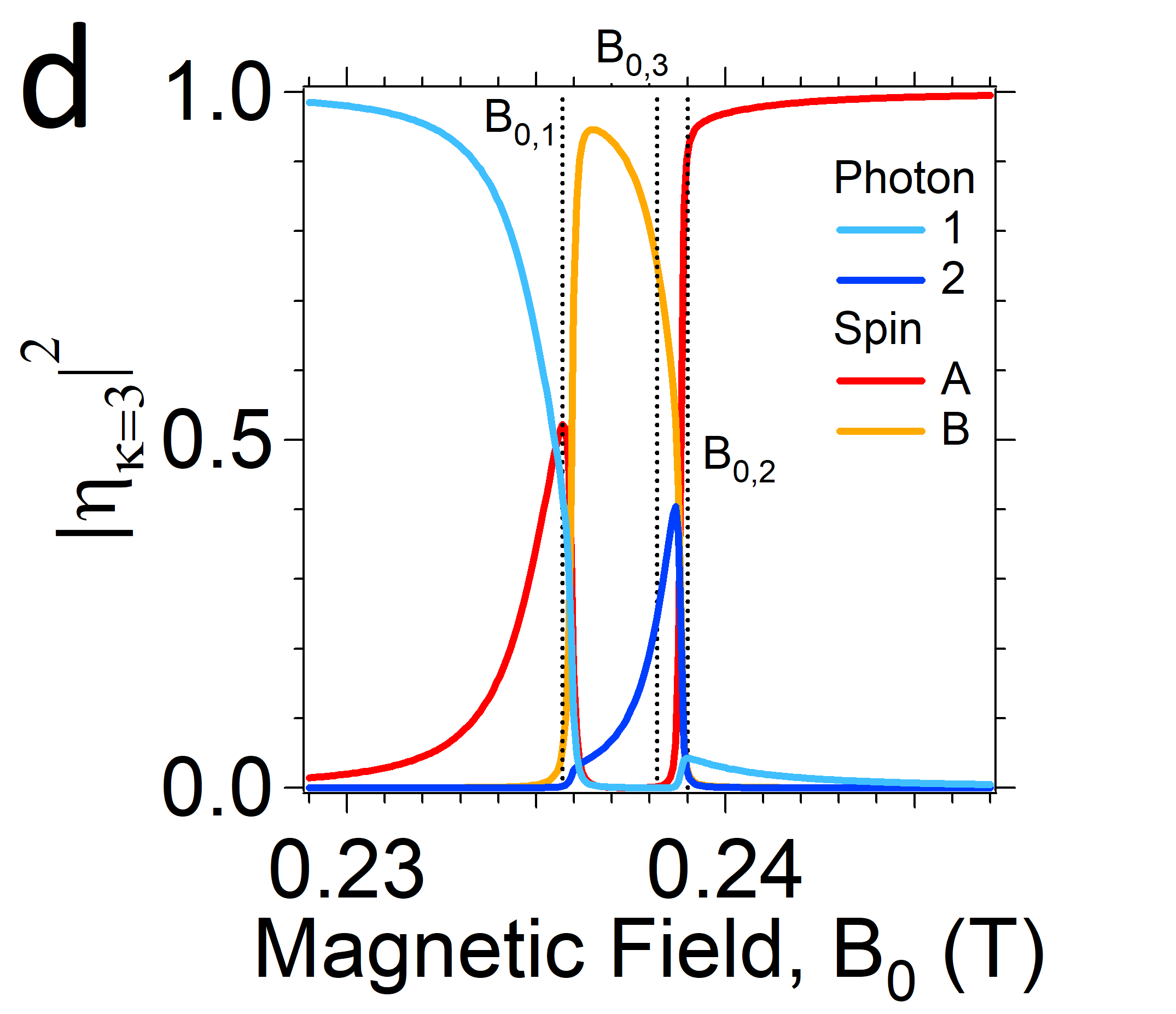}
\centering
\caption{\label{Fig04} Weight of the spin and photon modes in the hybridized system modes $k=2$ (a,c) and $k=3$ (b,d) calculated for the \#A+\#B case, both in the parallel (a,b) and in the perpendicular (c,d) configurations. Vertical dashed lines with labels show the positions of the resonant fields discussed in the text.}
\end{figure}

The couplings between cavity modes and multiple spin ensembles in principle allow for a high degree of coherent mixing in the system modes. Hereafter, we discuss such mixing in terms of the composition of the hybridized modes $c_k = \sum_{l=1,2} \eta_{k,l} a_{l} + \sum_{\chi=A,B} \eta_{k,\chi} b_{\chi} $ that diagonalize the system Hamiltonian. Fig. \ref{Fig04} shows the square moduli of the coefficients $\eta_{k,l}$,$\eta_{k,\chi}$ obtained as a function of the magnetic field from the set of parameters reported in Tables \ref{Table1} and \ref{Table2}. While in the case of a single spin ensemble coupled to a single resonator mode, the hybridized modes can only undergo a transition from photon to spin (see \cite{ESI,ghirriPRA2016}), or viceversa, here further possibilities show up. In particular, the mode $k=2$ undergoes a spin-spin transition (from \#A to \#B) at the first resonant field, and one from \#A to \#2 around the second resonant field, where in fact the two spin modes hybridize with the photon mode \#2 [panel (a)]. An analogous behavior is obtained for the mode $k=3$ (b). Conversely, the hybrid modes $k=1,4$ undergo a simple one spin-one photon mode transition at the corresponding resonant fields (see \cite{ESI}). Considering now a DMR with a smaller energy difference between the two resonator modes (c,d), the transitions occurring at the two anticrossings partially overlap and the  three-mode mixing occurs at both the resonant fields. Furthermore the system approaches a condition in which a small mixing between the two photon modes occurs due to the presence of the spin ensembles. 
These kind of interplay involving the photon modes are also visible in the parallel geometry (Fig. \ref{Fig04}.a). In fact, the mode $k=2$ displays a transition between the two spin components (ensemble \#A and \#B) around $B_{0}=B_{0,1}$, where also a small contribution of photon mode \#1 is also present. Slightly below the resonant field $B_{0,2}$, a new transition between the two spin modes takes place, in which photon mode \#2 is now giving a small contribution. Finally, at the second resonant field, the transition occurs between the spin component corresponding to \#B and the photon mode \#2. Here, although being weakly coupled to Mode \#1, ensemble \#A has a non negligible effect. Similar considerations hold also for the hybridized mode $k=3$ (Fig. \ref{Fig04}.b), for which the role and the interplay of the spin and photon components as a function of the magnetic field is specular.\\ Notably a high degree of spin mixing is reached also for magnetic fields in the range between the two resonant fields of the DMRs in the parallel geometry. Conversely, in the perpendicular geometry the frequency difference between mode \#1 and \#2 becomes comparable to the width of the avoided level crossings. This reflects also in the composition of the hybridized modes $k=2,3$ (Fig. \ref{Fig04}.c,d), which still show the specular interplay between two spin and one photonic components. The main difference with respect to the previous case is given by the amount of mixing around the crossing fields. In fact, the two consecutive transitions close to the higher (lower) magnetic field values for $k=2$ ($k=3$) collapse now at the $B_{0,2}$ ($B_{0,1}$) field.

\subsection{\label{sec.entropy_spin_char}Modal Entropy and Spin Character}

\begin{figure}[tb]
\centering
\includegraphics[width=0.47\columnwidth]{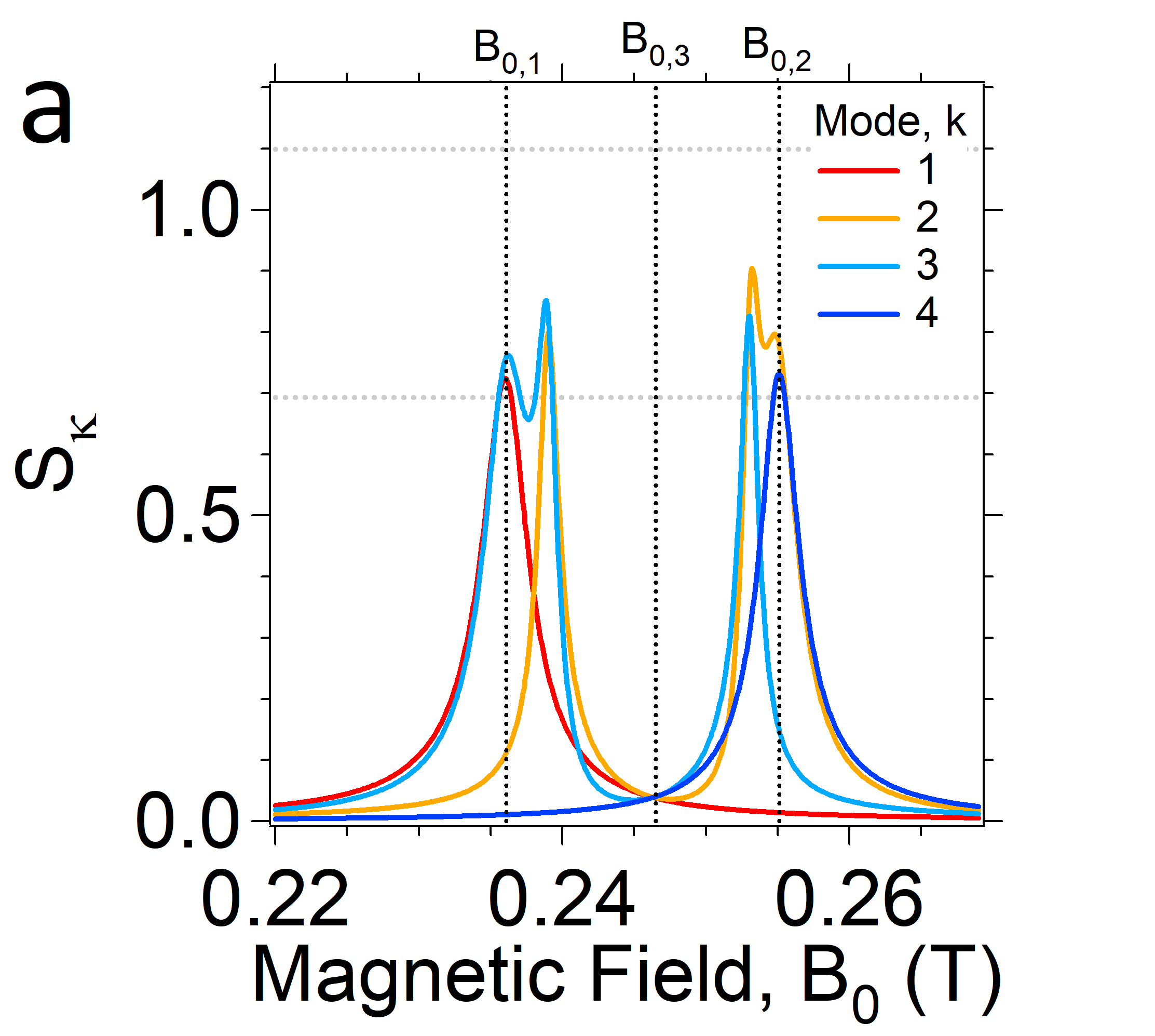} 
\includegraphics[width=0.50\columnwidth]{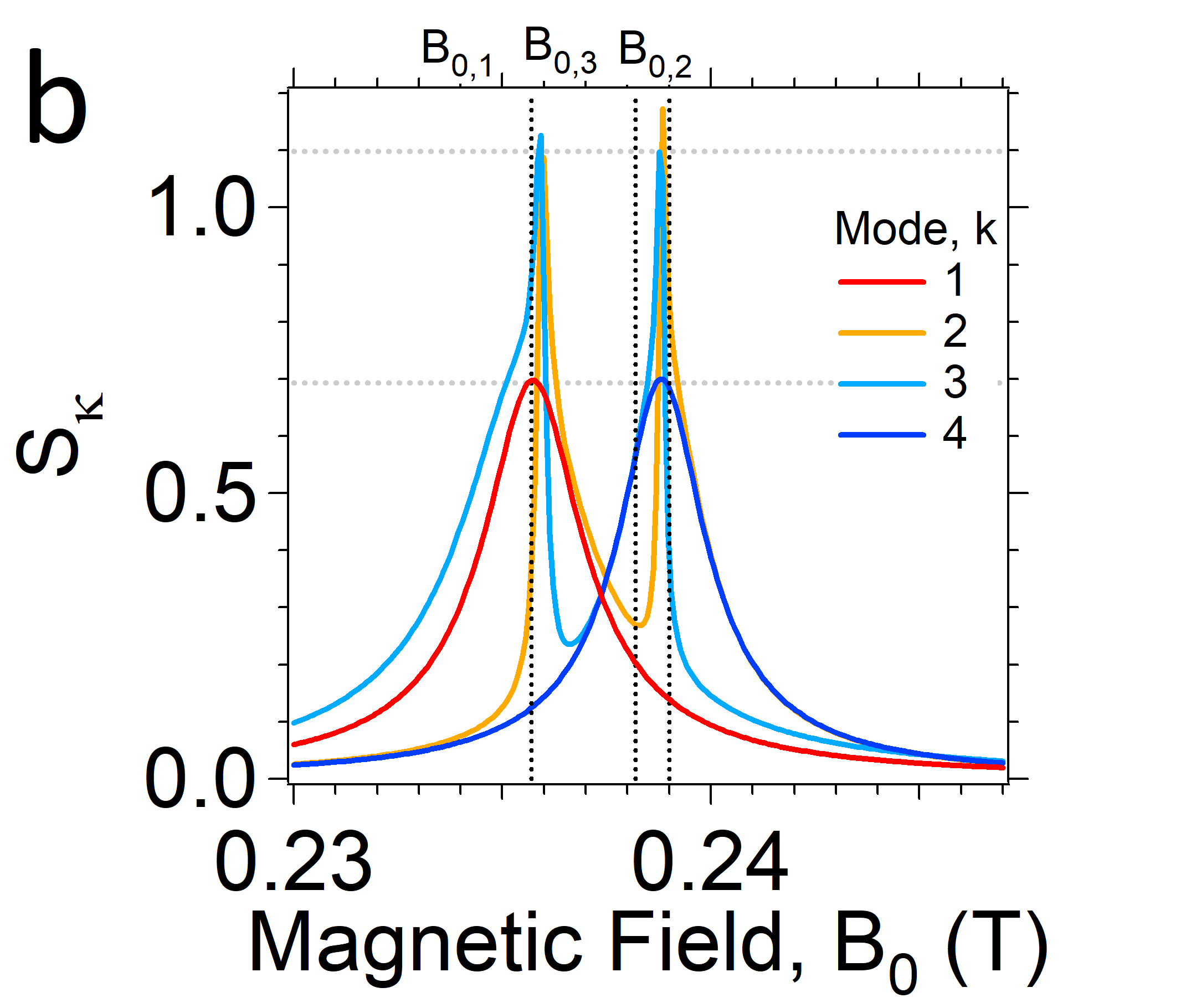} 
\includegraphics[width=0.48\columnwidth]{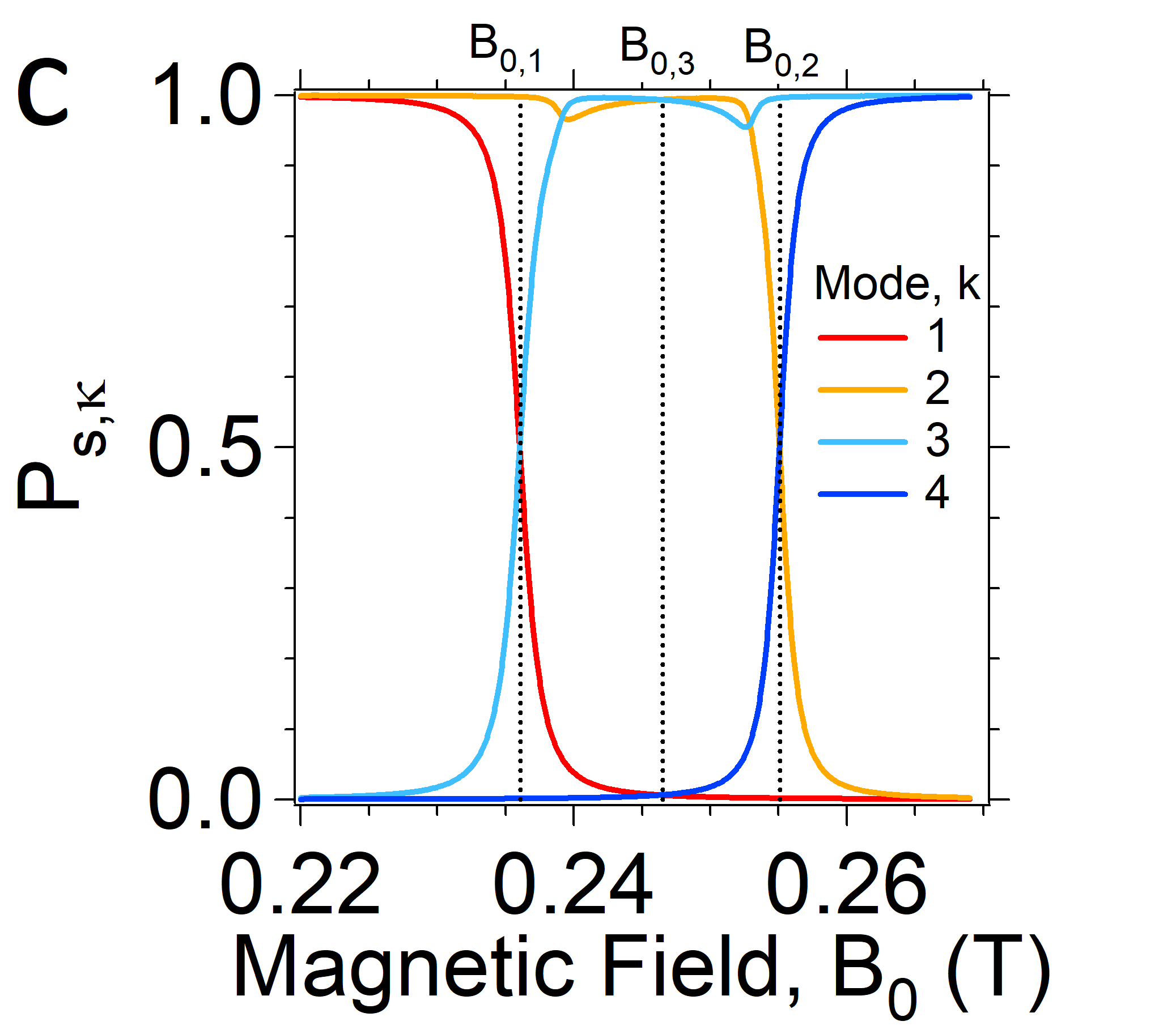} 
\includegraphics[width=0.50\columnwidth]{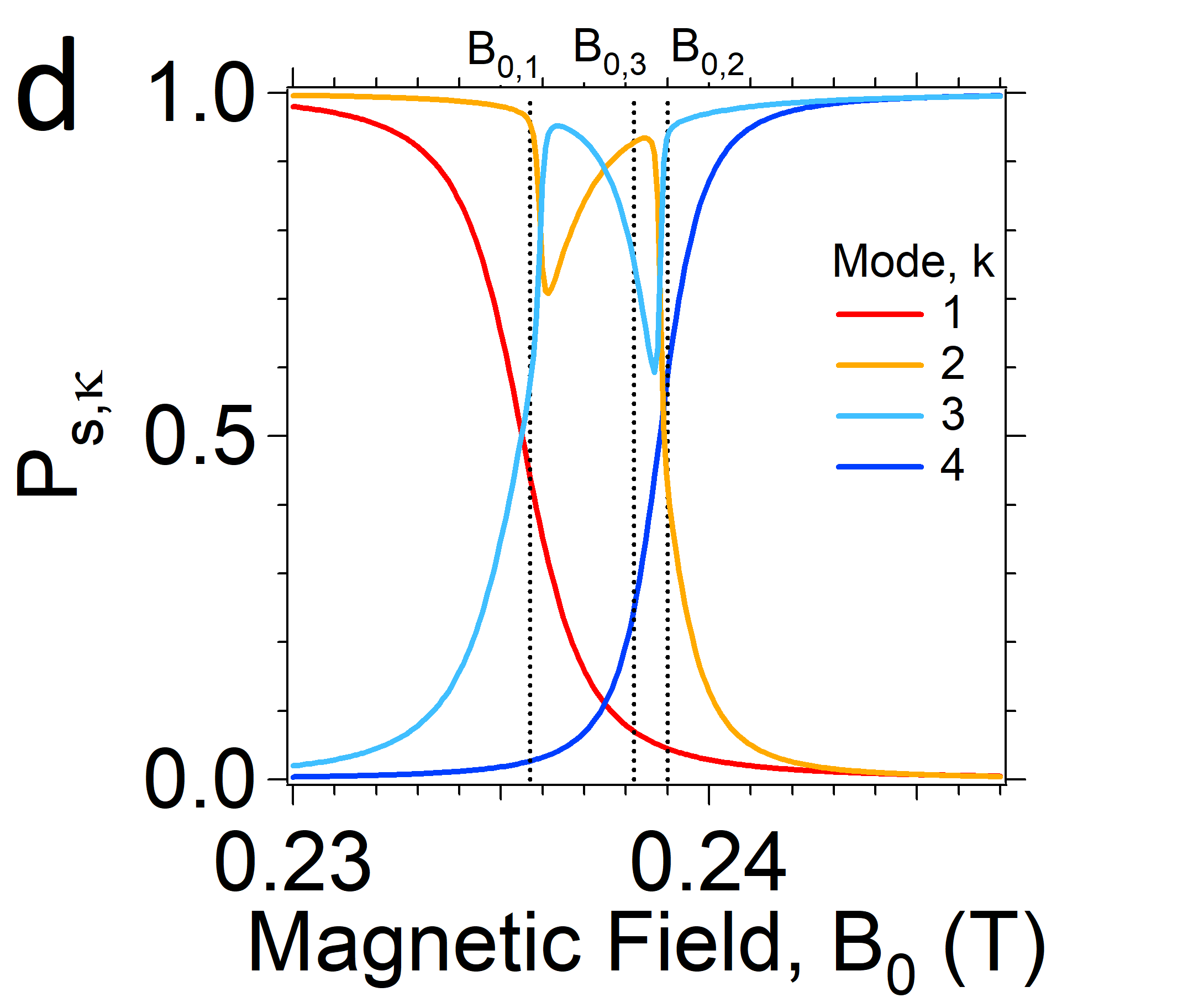} 
\centering
\caption{Characterization of the hybridized modes in term of their modal entropy $S_k$ (a,b) and the spin character $P_{s,k}$ (c,d) as a function of the static magnetic field. The case of the \#A + \#B sample positioning discussed in the main paper (Fig. 2 and Fig. 3) are considered. Here (a,c) refer to the DMR with parallel coupling, while (b,d) refer to perpendicular one. Horizontal dotted lines in (a,b) correspond to $ln\,2$ and to $ln\,3$, respectively.}
\label{fig.entropy_spin_char}
\end{figure}

Fig. \ref{fig.entropy_spin_char} gives the Modal Entropy (Eq. \ref{eq.modal_entropy}) and the Spin Character (Eq. \ref{eq.spin_character}) calculated as a function of the magnetic field according to the amplitudes given in Fig. \ref{Fig04}. The presence of coherent mixing between spin and photon modes is denoted by the peaks in the modal entropy $S_k$. In the parallel coupled DMR a doublet of peaks in the modal entropy is visible at each of the resonant fields ($B_{0,1},B_{0,2}$). For the hybrid modes $k=1$ and $k=4$ the entropy is slightly above the $ln\,2$ value, confirming the mixing is essentially involving two bare components and only a small contribution from a third one. The spin character $P_{s,k}$ confirm this single ensemble-single photon mixing, since the hybrid modes $k=1,4$ undergo a transition from 1 to 0 and from 0 to 1 at their two corresponding resonant fields (see \cite{ESI}). Conversely, the hybrid modes $k=2,3$ reach values of modal entropy higher than $ln\,2$, confirming that the mixing involves three bare components. The spin character is always around 1 between the two resonant fields, suggesting that the hybridization is effectively mediated by the spin ensembles.

In the perpendicular coupling case the main differences are visible for modes with $k=2$ and $k=3$, consistently to what reported in Fig. \ref{Fig04} (see also Fig. S.6 and S.7 of \cite{ESI}) The modal entropy shows now shows two sharp peaks in correspondence of the two resonant fields, and its values are always well above $ln\,2$. One of the two peaks slightly goes above $ln\,3$, corroborating the fact that the system is approaching a condition for a four mode hybridization. The spin characters for $k=2,3$ now show and additional dip as a function of the static magnetic field and always remain above 50 \% between the two resonant fields. This again corroborates the dominant spin character of the mode hybridization and that the photon mixing is effectively mediated by the spins.

\section{Conclusions}

In conclusion, we have investigated YBCO dual mode resonators for circuit QED. Both the resonant frequencies and the phases of the coupling to external field modes can be widely tuned by means of the device geometry. The coherent coupling between the spin ensembles is achieved for both the cavity modes. It results in a coherent mixing of the system modes, involving two spatially separated spin ensembles and at least one photon mode. A small - albeit non negligible - mixing between the two photon modes, mediated by the spin ensembles, also emerges. The results are corroborated also by entropy measures performed on the system. A possible way to enhance this effect might be a further reduction of the asymmetry between the two resonant modes. We finally mention that DMR offer further additional possibilities in the perspective of tailoring the mode hybridization thanks to the intrinsic non linearities resulting from the kinetic inductance of the superconductor \cite{monacoJAP2000,dahmAPL1996}.

\begin{acknowledgments}
We thank Prof. A. Cassinese (UniNA) for useful discussions. This work is partially funded by the Italian Ministry of Education and Research (MIUR) through PRIN Project (contract no 2015HYFSRT) and by the Air Force Office of Scientific Research grant (contract no FA2386-17-1-4040).\\

The authors declare no competitive financial interest.
\end{acknowledgments}


\bibliographystyle{unsrt}
\bibliography{biblio}

\begin{thebibliography}{10}

\bibitem{walls}
D.~F. Walls and Gerard~J. Milburn.
\newblock {\em Quantum Optics}.
\newblock Springer, 2007.

\bibitem{ghirriPRA2016}
A.~Ghirri, C.~Bonizzoni, F.~Troiani, N.~Buccheri, L.~Beverina, A.~Cassinese,
  and M.~Affronte.
\newblock Coherently coupling distinct spin ensembles through a high-${T}_{c}$
  superconducting resonator.
\newblock {\em Phys. Rev. A}, 93:063855, Jun 2016.

\bibitem{bonizzoniAdvPhys2018}
C.~Bonizzoni, A.~Ghirri, and M.~Affronte.
\newblock Coherent coupling of molecular spins with microwave photons in planar
  superconducting resonators.
\newblock {\em Advances in Physics: X}, 3(1):1435305, 2018.

\bibitem{zhuIEEEtransMicrowTech1999}
L.~Zhu, P.~M. Wecowski, and K.~Wu.
\newblock New planar dual-mode filter using cross-slotted patch resonator for
  simultaneous size and loss reduction.
\newblock {\em IEEE Transactions on Microwave Theory and Techniques},
  47(5):650--654, May 1999.

\bibitem{najiSCIREP2017}
Adham Naji and Mina~H. Soliman.
\newblock Center frequency stabilization in planar dual-mode resonators during
  mode-splitting control.
\newblock {\em Scientific Reports}, 7:43855--, March 2017.

\bibitem{kanayaASIEEE2003}
H.~Kanaya, J.~Fujiyama, R.~Oba, and K.~Yoshida.
\newblock Design method of miniaturized hts coplanar waveguide bandpass filters
  using cross coupling.
\newblock {\em Applied Superconductivity, IEEE Transactions on},
  13(2):265--268, June 2003.

\bibitem{bourgeoisIEEEMicrTech2005}
P.~Y. Bourgeois and V.~Giordano.
\newblock Simple model for the mode-splitting effect in whispering-gallery-mode
  resonators.
\newblock {\em IEEE Transactions on Microwave Theory and Techniques},
  53(10):3185--3190, Oct 2005.

\bibitem{ghirriAPL2015_notce}
A.~Ghirri, C.~Bonizzoni, D.~Gerace, S.~Sanna, A.~Cassinese, and M.~Affronte.
\newblock $yba_{2}cu_{3}o_{7}$ microwave resonators for strong collective
  coupling with spin ensembles.
\newblock {\em Applied Physics Letters}, 106(18):184101, 2015.

\bibitem{pozar}
David~M. Pozar.
\newblock {\em Microwave Engineering}.
\newblock John Wiley \& Sons, 4th edition edition, 2012.

\bibitem{sageJAP2011}
Jeremy~M. Sage, Vladimir Bolkhovsky, William~D. Oliver, Benjamin Turek, and
  Paul~B. Welander.
\newblock Study of loss in superconducting coplanar waveguide resonators.
\newblock {\em Journal of Applied Physics}, 109(6):063915, 2011.

\bibitem{chiorescuPRB2010}
I.~Chiorescu, N.~Groll, S.~Bertaina, T.~Mori, and S.~Miyashita.
\newblock Magnetic strong coupling in a spin-photon system and transition to
  classical regime.
\newblock {\em Phys. Rev. B}, 82:024413, 2010.

\bibitem{holsteinPhysRev1940}
T.~Holstein and H.~Primakoff.
\newblock Field dependence of the intrinsic domain magnetization of a
  ferromagnet.
\newblock {\em Phys. Rev.}, 58:1098--1113, Dec 1940.

\bibitem{abeAPL2011}
Eisuke Abe, Hua Wu, Arzhang Ardavan, and John J.~L. Morton.
\newblock Electron spin ensemble strongly coupled to a three-dimensional
  microwave cavity.
\newblock {\em Applied Physics Letters}, 98(25):251108, 2011.

\bibitem{jenkinsJOP2013}
Mark Jenkins, Thomas H\"ummer, Mar\'ia~Jos\'e Mart\'inez-P\'erez, Juanjo
  Garc\'ia-Ripoll, David Zueco, and Fernando Luis.
\newblock Coupling single-molecule magnets to quantum circuits.
\newblock {\em New Journal of Physics}, 15:095007, 2013.

\bibitem{bonizzoniSCIREP2017}
C.~Bonizzoni, A.~Ghirri, M.~Atzori, L.~Sorace, R.~Sessoli, and M.~Affronte.
\newblock Coherent coupling between vanadyl phthalocyanine spin ensemble and
  microwave photons: towards integration of molecular spin qubits into quantum
  circuits.
\newblock {\em Scientific Reports}, 7(1):13096, 2017.

\end{thebibliography}

\end{document}